\documentclass[a4paper,11pt]{article}
\pdfoutput=1

\usepackage[utf8]{inputenc}
\usepackage[T1]{fontenc}
\usepackage[english]{babel}

\usepackage{indentfirst}
\usepackage{amsmath}
\usepackage{amsthm}
\usepackage{amssymb}
\usepackage{graphicx}
\usepackage{graphics}
\usepackage{psfrag}
\usepackage[font=small]{caption}
\usepackage{subcaption}
\usepackage{color}
\usepackage{pgf,pgfnodes}
\usepackage{bbold}

\allowhyphens

\newcommand{\ketB}[1]{{\left|#1\right\rangle}_{\text{B}}}
\newcommand{\ketF}[1]{{\left|#1\right\rangle}_{\text{F}}}

\newcommand{\ordo}{\mathcal{O}}
\newcommand{\lecho}{\mathcal{L}(t)}

\newcommand{\ket}[1]{{\left|#1\right\rangle}}
\newcommand{\bra}[1]{{\left\langle #1\right|}}

\newcommand{\skalarszorzat}[2]{{\langle #1 | #2 \rangle}}

\newcommand{\vev}[1]{\left\langle #1 \right\rangle}

\newcommand{\identity}{\mathbb{1}}

\usepackage{ifpdf}

\ifpdf
\usepackage{epstopdf}
\usepackage[pdftex,colorlinks,urlcolor=blue,citecolor=blue,linkcolor=blue]{hyperref}
\else
\usepackage[hypertex,colorlinks,urlcolor=blue,citecolor=blue,linkcolor=blue]{hyperref}
\fi

\makeatother

\begin{document}

\numberwithin{equation}{section}

\title{Real-time dynamics in a strongly interacting bosonic hopping
  model: Global quenches and mapping to the XX chain}
\author{Bal\'azs Pozsgay$^1$ and Viktor Eisler$^{2,3}$
\\
~\\
 $^{1}$MTA--BME \textquotedbl{}Momentum\textquotedbl{} Statistical
Field Theory Research Group\\
Budafoki \'ut 8, H-1111 Budapest, Hungary\\
$^2$Institut f\"ur Theoretische Physik, Technische Universit\"at Graz, \\
Petersgasse 16, A-8010 Graz, Austria \\
$^3$MTA-ELTE Theoretical Physics Research Group,
E\"otv\"os Lor\'and University,\\
 P\'azm\'any P\'eter s\'et\'any 1/a, H-1117 Budapest, Hungary
}
\maketitle

\abstract{
We study the time evolution of an integrable many-particle system,
described by the $q$-boson Hamiltonian in the limit of strong bosonic interactions $q\to\infty$.
It is shown that, for a particular class of pure initial states, the analytical calculation
of certain observables simplifies considerably. Namely, we provide exact
formulas for the calculation of the Loschmidt-echo and the emptiness formation
probability, where the computational time scales polynomially with the particle number.
Moreover, we construct a non-local mapping of the $q$-boson model to the
XX spin chain, and show how this can be utilized to obtain the time evolution
of various local bosonic observables for translationally invariant initial states.
The results obtained via the bosonic and fermionic picture show
perfect agreement. 
In the infinite volume and large time limits, we rigorously verify the
prediction of the Generalized Gibbs Ensemble for homogeneous initial
Fock states.}

\section{Introduction}

The study of time evolution in isolated quantum many-body systems has become
a leading direction of research within both experimental and theoretical condensed
matter physics
\cite{Silva-quench-colloquium,QM-out-of-equilibrium-review}. One of
the main questions to be answered is whether unitary time 
evolution can lead to a local thermalization of the system and what are the
characteristics of the relaxation towards this stationary state.
Particularly interesting is the case of integrable quantum systems where the dynamics
is severely constrained by an extensive set of conservation laws. In order to take
these constraints into account, an extended notion of thermalization is required
and a statistical ensemble under the name of Generalized Gibbs Ensemble (GGE)
was put forward for the description of the stationary state \cite{rigol-gge}.

One way of testing the validity of the GGE is by monitoring the asymptotic time evolution of local
observables, which is a notoriously hard task for genuinely interacting
Hamiltonians. One possibility is to resort to purely numerical techniques
such as the time evolving block decimation method \cite{vidal-itebd1,vidal-itebd2} applied in
\cite{sajat-oTBA}, or to develop special tools based on
integrability. From the analytical side, one of the most successful approaches to give predictions for the long
time limit of observables is the so-called quench action technique \cite{quench-action}.
It provides a means of capturing the stationary state emerging from the unitary time
evolution of a Bethe Ansatz integrable model from a pure initial state.
The method has been successfully applied for some simple initial states of
the 1D Bose gas \cite{caux-stb-LL-BEC-quench} and the infinite XXZ
chain \cite{JS-oTBA,sajat-oTBA}. Remarkably, in the latter case it has been 
found that the stationary state can only be described 
by a GGE that is supplemented by a new family of quasi-local conserved
charges \cite{JS-CGGE}. Recently, some generalizations to finite size chains have also been
reported \cite{finite-qa}.

Despite the success of the Quench Action method, its scope is mainly
restricted to the asymptotic regime of the time evolution. Indeed,
the technique relies on selecting a representative Bethe state of the interacting
Hamiltonian via a saddle point analysis, which in turn encodes all the information
about the time-evolved state for $t \to \infty$. 
However, if one is interested
in early or intermediate time-scales, the summation over the complete set
of Bethe states, contributing to the time evolution of a specific observable,
can not be avoided. In order to overcome the limitations of summing over an
exponentially large set of eigenstates, one needs some prior knowledge
about the importance of the contributions from many-particle states,
which is a crucial ingredient behind Bethe Ansatz based numerical methods \cite{ABACUS,JS-Milos}.
Although such an importance sampling is feasible for some integrable
models \cite{JS-asympt-from-QA},
it is not yet clear whether strict  analytical results could be
obtained using this technique.

On the other hand, within the field of Bethe Ansatz integrable systems,
there has always been an immense theoretical interest in devising exact
analytical techniques. Indeed, several successful approaches exist for the
computation of equilibrium properties of these models, in particular the
Heisenberg XXZ chain and the 1D Bose gas \cite{korepin-Book}. 
Important achievements include the computation of the asymptotics of
space and time dependent correlation functions (see \cite{karol-hab}
and references therein) and the derivation of effective, factorized
formulas for the  ground state and finite temperature local
correlators of the XXZ model (see \cite{kluemper-goehmann-finiteT-review} and
references therein). In view of these remarkable achievements the question naturally arises whether some
progress can be made for the far-from-equilibrium physics of
integrable systems. Such studies are also motivated by the fact that
in certain quenches of the XXZ chain and the 1D Bose gas exact results have been obtained
for the stationary states 
\cite{caux-stb-LL-BEC-quench,JS-oTBA,js-hosszu-kvencs,sajat-QA-GGE-hosszu-cikk,JS-CGGE,jacopo-michael-hirota}.
Ideally, one would like to have exact formulas for the full time
evolution, such that the previously mentioned results could be
obtained simply by taking the long time limit. Whether or not such
program can be carried out is not clear at the moment. 

Motivated by these long term goals, in the present paper we set a
somewhat simpler objective: we consider time evolution in the
$q$-boson Hamiltonian in the $q \to \infty$ limit, 
also known as the phase model
\cite{qboson-izergin-kitanine-bog,qboson-bog1,qboson-bog2,qboson-bog3}. 
In this limit the bosonic interactions are strong, 
there is non-trivial scattering between the particles, but the
scattering is simple enough so that manageable exact expressions can be
obtained for the observables.
Therefore, the complexity of this lattice 
hopping model is somewhere between that of free theories and a generic
Bethe Ansatz solvable model.
The quantities of interest are the return probability (or Loschmidt echo) and the
emptiness formation probability (EFP) of the time evolved
state. 
The EFP is probably the simplest local observable for which efficient
closed-form expressions can be found for integrable models \cite{KIEU94,EFIK95}.  
On the other hand, the calculation of
  the Loschmidt echo mainly serves as an introductory example to
  demonstrate our method. Namely, we show
  that, due to the simple constraints between rapidities 
and for a specific class of initial states, the sums over exponentially many
Bethe states can be turned into a simple sum over the total momentum and an
auxiliary variable. Hence, we obtain an exact analytical expression for the
Loschmidt echo where the number of terms to be summed scales at most
linearly in both the system size and the number of particles.
Interestingly, it turns out that the form factors of the
EFP have again the properties which allow the same trick to be carried out,
i.e. the exponential sums over intermediate states can again be replaced
by an expression in which the number of terms scales polynomially.

The huge simplification in the computational efforts of the above quantities
suggests that there might be some deeper connection between the $q$-boson
Hamiltonian and a non-interacting model. In fact, such a mapping to the XX spin chain
(which is equivalent to a free-fermion hopping model) was already pointed out in an earlier
work \cite{qboson-bog3}. Here we show that, although some boundary terms spoil an exact correspondence,
the mapping can be symmetrized such that the two models become equivalent in the
zero-momentum sector even for finite chain sizes. Furthermore, even though the mapping is
non-local, we show that the average EFP's are exactly mapped onto each other.
Thus, for translationally invariant bosonic initial states, the calculation of the time
evolution of the EFP further simplifies using the free-fermion
representation. Moreover, the fermionic methods allow us to compute
other observables (such as the local bosonic occupation probabilities)
that were previously inaccessible with the bosonic
approach. 

In the following, we first introduce the $q$-boson model in section 2 and
describe its Bethe Ansatz solution in the limit $q=\infty$. In section 3
we report our main results about the analytical calculation of the time evolution
of the Loschmidt echo and the EFP for a simple class of initial states.
Sec. 4 describes the non-local mapping from the $q$-boson Hamiltonian to
the XX chain and introduces the free-fermion formalism for the calculation of the EFP.
Some particular translation invariant states are considered in Sec. 5
where the numerical results on the EFP from our new Bethe Ansatz approach
are cross-checked to analytical formulas obtained via the fermionic representation.
Our concluding remarks are found in Sec. 6. The paper is supplemented
with two appendices where some details of the analytical calculations are given.

\section{The model and its Bethe Ansatz solution}

Consider a bosonic chain of length $L$. The Hilbert space is spanned
by Fock states
\begin{equation}
\label{Fock}
  \ket{n_1,n_2,\dots,n_L}=\ket{n_1}_1\otimes\ket{n_2}_2\otimes\dots
  \otimes \ket{n_L}_L,
\end{equation}
where $n_j\ge 0$ represent the local occupation numbers. 

The $q$-boson model at $q=\infty$ (also called the phase model) is
given by the Hamiltonian \footnote{In the present work we do not treat
the model for general $q$. The reader who is interested in the general
case is referred to \cite{sajat-qboson} and references therein.}
\begin{equation}
  \label{Hqq}
H_{\text{B}}=-\sum_{j=1}^{L} (\phi_j^\dagger \phi_{j+1}+ \phi_{j+1}^\dagger \phi_{j}-2N_j),
\end{equation}
where the operators $\phi_j$, $\phi^\dagger_j$ are defined by their action 
\begin{equation}
\label{fik}
  \phi_j\ket{n}_j=
  \begin{cases}
\ket{n-1}_j, & \text{for } n>0 \\
0, & \text{for } n=0,\\    
  \end{cases}\qquad\text{and}
\qquad \phi^\dagger_j \ket{n}_j=\ket{n+1}_j
\end{equation}
on local bosonic states
and the $N_j$ are the standard local number operators. In \eqref{Hqq}
periodic boundary conditions are assumed.
Note that the
$\phi$, $\phi^\dagger$ operators do not coincide with the standard
bosonic creation/annihilation operators, in particular they satisfy
the somewhat unusual exchange relation
\begin{equation*}
  [\phi_j,\phi_k^\dagger]=\delta_{j,k} \delta_{n_j,0}.
\end{equation*}
Although there is no explicit interaction term in the Hamiltonian
\eqref{Hqq}, it is not free: due to definition \eqref{fik} the
physical hopping
amplitudes depend on the local occupation numbers.

The $q$-boson model can be solved by the different versions of the Bethe
Ansatz. The Algebraic Bethe Ansatz (ABA) solution was first derived (for
general $q$) in the papers 
\cite{qbozon-bog-bullough1,qbozon-bog-bullough2,qbozon-bog-bullough3},
whereas equilibrium correlation functions were calculated in
\cite{qboson-izergin-kitanine-bog}. Afterwards, the coordinate Bethe Ansatz wave
functions were given in \cite{qboson-bog1,qboson-bog3}. The phase
model is
also related to the enumeration of plane partitions
\cite{qboson-bog1,qboson-bog2,qboson-bog3,qboson-keiichi}.

In the present
work we refrain from discussing the  ABA solution as all our
results can be obtained from the known real space wave
functions. 
We write the $N$-particle eigenstates of the system 
as
\begin{equation}
\label{BetheStateQ}
 \ket{\{a\}_N}
=
\sum_{1\le x_1\le x_2\le \dots x_N\le L} 
 C_N(x_1,\dots,x_N)  
 \phi^\dagger_{x_1}\dots \phi^\dagger_{x_N}\ketB{0}.
\end{equation}
It was derived in \cite{qboson-bog1} that the coefficients can be
expressed as
\begin{equation}
\label{coeff}
 C_N(x_1,\dots,x_N)=  
\det_N\Big( (a_k)^{j+x_j}\Big),
\qquad a_k=e^{ip_k},
\end{equation}
where the $p_k$ variables in \eqref{coeff} can be identified as quasi-momenta of the
interacting particles.
We note that our formula \eqref{coeff} differs from the
conventions of the ABA literature in both the overall norm and the
phase of the vector. In particular, there is a sign depending on the ordering
of the parameters $a_j$, however this drops out from the actual
calculations. 
Also, we will assume for simplicity that $N$ and $L$ are
even. Odd values only affect certain signs in intermediate results,
but not the physical observables.

In a Bethe Ansatz wave function a simple phase arises when we exchange
two particle positions, and this is interpreted as the
physical $S$-matrix of the particles. In the present case we can read
off \eqref{coeff} that this phase is 
\begin{equation}
\label{S}
  S(p_1,p_2)=-e^{i(p_1-p_2)}.
\end{equation}
Periodicity of the wave function implies the Bethe equations:
\begin{equation*}
  e^{ip_jL}\prod_{k\ne j}S(p_j-p_k)=1.
\end{equation*}
In the phase model this gives
\begin{equation}
\label{BeQ}
  (a_j)^{L+N} =(-1)^{N-1}e^{iP},\qquad e^{iP}=\prod_{k=1}^N a_k,
\end{equation}
where $e^{iP}$ is the eigenvalue of the one-shift translation
operator. Equation \eqref{BeQ} shows that the quantization condition is ``almost
free'': once $P$ is fixed the momenta can be chosen independently, as long as they
satisfy the overall constraint on the r.h.s. of \eqref{BeQ}. However,
the exponent of the variable $a_j$ is $L+N$, whereas a free theory
would give simply $L$.

If the BA equations are satisfied, the energy eigenvalues are given by
\begin{equation}
\label{enmom}
  E_N=\sum_{j=1}^N e(p_j)\quad\text{where}\quad e(p)=4\sin^2(p/2).
\end{equation}
The norm of the Bethe states is simply
\begin{equation}
\label{normQ}
  \skalarszorzat{\{a\}_N}{\{a\}_N}
=L(L+N)^{(N-1)}. 
\end{equation}
This result was obtained in \cite{qboson-izergin-kitanine-bog} using
the Algebraic Bethe Ansatz, whereas in \cite{qboson-bog1} it was
shown that it follows from certain properties of the Schur
polynomials \footnote{In
  \cite{qboson-izergin-kitanine-bog,qboson-bog1} the norm formula
  includes a Vandermonde determinant, which arises due to the
  different normalization of the Bethe vectors.}. 

In the thermodynamic limit $N,L\to \infty$ and $N/L=n$ it is convenient to work
with root and hole densities $\rho_r(p)$ and $\rho_h(p)$. In generic
Bethe Ansatz solvable models these densities satisfy linear integral
equations, however in the present case the constraint for them reads simply
\begin{equation}
\label{rhoegyinfty}
  \rho_r(p)+\rho_h(p)=1+n,
\end{equation}
which can be obtained directly from the Bethe equations \eqref{BeQ}.

\section{Finite volume real time dynamics}

\label{sec:exact}

In this section we present a new numerical method to study
real time dynamics in the phase model. Our method is
similar to that of \cite{qboson-izergin-kitanine-bog}, where
two-point functions were computed in equilibrium.

In all cases below the setting is the following. At $t=0$
the system of finite volume $L$ is prepared in a state $\ket{\Psi_0}$, which can be a
simple Fock state in the coordinate basis, or some other state
prepared according to certain rules. At times $t>0$ the system is
subject to time evolution governed by the Hamiltonian \eqref{Hqq} and our goal is to
compute the physical observables.

We consider a class of initial states where the unnormalized overlaps can be written as
\begin{equation}
\label{talanilyen}
  \skalarszorzat{\Psi_0}{\{a\}_N}=
Z(N) \det G_j(a_k),
\end{equation}
where $Z(N)$ is a numerical constant, and $G$ is an $N\times N$ matrix
where the $k$th column only depends on the parameter $a_k$. It is easy to
see that pure Fock states in the coordinate basis satisfy this requirement:
 the overlap can be read off from \eqref{coeff} and we get
\begin{equation}
\label{FockG}
  G_j(a_k)=a_k^{x_j+j}, \qquad Z(N)=1.
\end{equation}
In the calculations below we will keep $G$ unspecified as long as
possible, but in the actual examples we will deal with $G$-matrices of the form \eqref{FockG}.

\subsection{Exact result for the Loschmidt-echo}

As a warm up we calculate the Loschmidt amplitude (or fidelity), which is defined as
\begin{equation}
  \lecho\equiv\skalarszorzat{\Psi_0}{\Psi(t)}=
\bra{\Psi_0}e^{-iH_Bt}\ket{\Psi_0}.
\end{equation}
Inserting a complete set of Bethe states and using the assumed form
\eqref{talanilyen}
we have
\begin{equation*}
  \lecho=\frac{Z^2(N)}{L(L+N)^{N-1}}\sum_{\{a\}}  |\det G_j(a_k) |^2 e^{-iE({a})t}.
\end{equation*}
Our goal is to perform the summation over the $a$-variables without
explicitly using the functions $G_j(a)$.
First we give an explicit solution of the Bethe equations \eqref{BeQ}.
Overall momentum quantization gives
\begin{equation}
  P=J\frac{2\pi}{L},\qquad J=1,\dots,L,
\label{Psol}
\end{equation}
and the $a$-variables can then be parametrized as
\begin{equation}
\label{asol}
 a_j=e^{ip_j},\qquad p_j=I_j\frac{2\pi}{N+L}+ \frac{\pi +P}{N+L}
\qquad I_j\in \{1,\dots,N+L\},
\end{equation}
where the $I_j$ are interpreted as momentum quantum numbers. These
parameters can not be chosen completely freely, because they have to
satisfy a constraint which follows from
\begin{equation}
  \sum_j p_j =P \text{ mod } 2\pi.
\end{equation}
The central idea of our method is to introduce an auxiliary sum
replacing the constraint so that we can sum over the
quantum numbers independently. 
We introduce the sum
\begin{equation}
\frac{1}{A}  \sum_{\alpha=1}^{A} e^{i\alpha(\sum_j p_j -P)},
\end{equation}
where $A$ is an integer depending on $N$ and $L$.
If the momentum constraint is satisfied, then the
sum gives 1, whereas in all other cases
\begin{equation}
  \sum_{\alpha=1}^{A} e^{i\alpha(\sum_j p_j -P)}=
e^{i(\sum_j p_j -P)}\frac{e^{iA(\sum_j p_j -P)}-1}{e^{i(\sum_j p_j -P)}-1}.
\end{equation}
We need to choose $A$ so that 
the r.h.s. above is always zero.
Using the solution \eqref{Psol}-\eqref{asol}
we obtain the necessary and sufficient conditions (assuming $N$ is even)
\begin{equation}
  \frac{A}{N+L}\in \mathbb{Z},\qquad
\frac{AN}{L(N+L)}\in\mathbb{Z}\qquad \frac{A}{2L}\in\mathbb{Z}.
\end{equation}
If we assume to have a fixed particle density, then $A$ has to scale linearly with $N$.
For example if $n=N/L=1$ then we can choose $A=2N$ to satisfy the
conditions. However, depending on the symmetries of the initial state
in many cases $A$ can be set to a fixed small number. This will
be discussed at the end of this section.

Now we are in the position to sum over the $a$-variables independently:
\begin{equation}
    \lecho=\frac{Z(N)^2}{AL(L+N)^{N-1}}\sum_{J=1}^L\sum_{\alpha=1}^A  
 \frac{1}{N!} \sum_{a_1,\dots,a_N} 
e^{i\alpha(\sum_j p_j -P)}
e^{-i\sum_j E(p_j)t}
\left|\det G_j(a_k) \right|^2.
\end{equation}
We expand the determinants as
\begin{equation}
  \left|\det G_j(a_k) \right|^2=
\sum_{\sigma,\bar\sigma} (-1)^\sigma(-1)^{\bar\sigma} \prod_j 
G_{\sigma_j}(a_{k})G^*_{\bar\sigma_j}(a_{k}).
\end{equation}
Each $a$ variable appears exactly twice in a factorized form,
therefore the summations over them can be
performed independently. By renaming the $a$-variables and reordering the factors in the product we
obtain a single sum and a factor of $(N!)$. 
The resulting sum has the form of a determinant, we thus obtain
\begin{equation}
\label{Lalakul}
      \lecho=\frac{Z(N)^2}{AL(L+N)^{N-1} }\sum_{J=1}^L\sum_{\alpha=1}^A  
e^{-i\alpha P} \det  M,
\end{equation}
with
\begin{equation}
   M_{jk}=\sum_{I=1}^{N+L} 
e^{i\alpha p_I}e^{-i E(p_I)t}
G_j(a_I)G^*_k(a_I),
\end{equation}
where
\begin{equation}
\label{asol2}
\begin{split}
a_I&=e^{ip_I},\qquad  p_I=I\frac{2\pi}{N+L}+ \frac{\pi +P}{N+L},\qquad  P=J\frac{2\pi}{L},\\
\end{split}
\end{equation}
and the energy is given by $E(p_I)=4\sin^2(p_I/2)$. We stress that in
\eqref{asol} there are $N$  $a$-variables for each state and they correspond to the
physical rapidities of the individual states, whereas in \eqref{asol2}
the $a$-variables form a list of the $N+L$ possible particle momenta for each overall
momentum $P$. To distinguish the two different roles of rapidities we
use upper case indices for the non-physical rapidity variables. 

Formula \eqref{Lalakul} expresses the Loschmidt echo as a sum of
$LA$ determinants, where the summation variables are the
overall momentum quantum number $J$ and the auxiliary parameter
$\alpha$ . Each matrix entering this sum can be 
written as
\begin{equation}
\label{jobb}
  M=\tilde G \Lambda \tilde G ^\dagger,
\end{equation}
where $\tilde G$ is an $N\times (L+N)$ matrix with elements given by
\begin{equation}
  \tilde G_{jK}=G_j(a_K),
\end{equation}
and $\Lambda$ is an $(L+N)\times
(L+N)$ diagonal matrix with elements
\begin{equation}
  \Lambda_{IK}=\delta_{IK} e^{i\alpha p_K}e^{-i E(p_K)t}.
\end{equation}
Note that $\tilde G$ only depends on $J$, whereas $\Lambda$ depends on both
$\alpha$ and $J$. Our derivation of \eqref{Lalakul}-\eqref{jobb} can be
regarded as a backwards application of the Cauchy-Binet formula, which
expresses the determinant of a product of non-square matrices as a sum
over subsets of the rows.

Depending on the situation, formula \eqref{Lalakul} can be transformed
into more convenient representations. One way to simplify it is to
introduce the  Fourier transform 
operator over $N$ points:
\begin{equation}
\label{Vdef}
  V_{kl}=\frac{1}{\sqrt{N}}e^{-iq_kl}\qquad\qquad q_k=\frac{2\pi}{N} k+\frac{\pi+P}{N}.
\end{equation}
For later convenience we introduced a shift in the $q$-variables so that
\begin{equation}
(e^{iq_k})^{N} = -e^{iP},\qquad k=1,\dots,N.
\end{equation}
It can be verified that $V$ is a unitary operator.
Inserting $V$ and $V^\dagger$ into \eqref{jobb} the Loschmidt echo is written as
\begin{equation}
\label{Lalakul3}
\begin{split}
      \lecho&=
\frac{1}{AL(L+N)^{N-1} }\sum_{J=1}^L
\sum_{\alpha=1}^A  e^{-i\alpha P} 
\det \left( H \Lambda H^\dagger\right),
\end{split}
\end{equation}
where
\begin{equation}
\label{Hdef}
  H_{jK}=\frac{1}{\sqrt{N}}\sum_{m} e^{-iq_j m}  G_m(a_K).
\end{equation}
This form is particularly useful if the initial state is a pure local Fock
state with $G$ given by \eqref{FockG}, because in those cases we
obtain a Fourier-like sum:
\begin{equation}
  H_{jK}=
\frac{1}{\sqrt{N}}\sum_{m} e^{-iq_j m}  e^{ip_K(x_m+m)}.
\end{equation}
If the initial state has periodic structure, 
then the matrix
$H$ can only
have few non-vanishing matrix elements. This makes formula
\eqref{Lalakul3} very effective. Note that the $q$-variables form a
set of $N$ whereas the $p$-variables form a set of 
$N+L$, and they have different definitions \eqref{Vdef} and
\eqref{asol2}.

Depending on the initial state, many of the determinants
in \eqref{Lalakul3} can be equal, thus the parameter $A$ can be
set to a fixed small number. Examples include periodic pure Fock
states. In appendix \ref{sec:equivalence} it is shown 
that if $x_j$ are the initial positions of the particles in the Fock
state and a new Fock state constructed from the coordinates
$y_j=j+x_j$ has periodicity $\nu$, then it 
is enough to choose $A=\nu$.

\subsection{Exact results for the emptiness formation probability}

\label{sec:EFPt}

Here we compute the time evolution of a simple physical observable,
namely the $m$-site emptiness formation probability (EFP). 
The EFP has been intensively studied in the ground state of various integrable systems
such as the XXZ chain
\cite{KIEU94,EFIK95,AK02,KMST02,KMST02b,KLNS03,Kozlowski08,Cantini12},
free-fermion Hamiltonians \cite{STN01,FA05,Stephan14} as well as the
$q$-boson model
\cite{qboson-izergin-kitanine-bog,qboson-bog1,qboson-bog2,qboson-bog3}. Recently,
there has even been some progress on the 
time evolution of the EFP after a quench for free fermions \cite{NR15}.

We define local projection operators $\Pi_{j}^0$ that project to
the subspace with no particle on site $j$. 
Then the
position dependent $m$-site EFP is defined as
\begin{equation*}
  E_{m}(j)=\prod_{k=0}^{m-1} \Pi_{j+k}^0.
\end{equation*}
Inserting two complete sets of states the time evolution is computed
as
\begin{equation*}
  \vev{E_m(j,t)}=
\skalarszorzat{\Psi_0}{\{a\}_N} \
\bra{\{a\}_N}E_{m}(j)\ket{\{b\}_N}\
\skalarszorzat{\{b\}_N}{\Psi_0}\
e^{-i(E_b-E_a)t}.
\end{equation*}
Matrix elements of the EFP operators were first derived in
 \cite{qboson-izergin-kitanine-bog}  using the ABA, but they can be calculated from the
coordinate BA wave functions too \cite{qboson-bog3}. In our
normalization
the
matrix element between two un-normalized off-shell Bethe states
reads
\begin{equation}
\label{FF}
\begin{split}
  \bra{\{a\}_N}E_{m}(j)\ket{\{b\}_N}&=
e^{i(P_2-P_1)j} \det\left[
  \frac{1-\left(\frac{b_l}{a_k}\right)^{N+L-m-1}}{1-\frac{b_l}{a_k}}
\right],
\end{split}
\end{equation}
where 
\begin{equation*}
  e^{iP_1}=\prod_{j=1}^N a_j \qquad  e^{iP_2}=\prod_{j=1}^N b_j.
\end{equation*}
If two rapidities coincide then the corresponding matrix element has
to be evaluated using l'H\^opital's rule. 

The matrix on the r.h.s. of \eqref{FF} has the property that its $l$th row
(or $k$th column) depends on $b_l$ (or $a_k$), respectively. Together
with the similar form of the overlap matrices, this makes it possible
to perform essentially the same steps as in the previous section. The difference
is that here we are faced with 
a double sum over
intermediate states, leading to two auxiliary sums.
This yields the following final result:
\begin{equation}
\label{efpT}
     \vev{E_m(j,t)}=\frac{Z^2(N)}
{(AL(L+N)^{N-1})^2}
\sum_{J_1,J_2=1}^{L}
e^{i(P_2-P_1)j}
\sum_{\alpha_1,\alpha_2=1}^{A}
 e^{-i\alpha_1 P_1} e^{i\alpha_2 P_2} 
\det O_m.
\end{equation}
Here $J_{1,2}$ are the momentum quantum numbers, $\alpha_{1,2}$
are the auxiliary variables, and the $N\times N$ matrix $O_m$ is given by
\begin{equation}
\label{om}
  O_m=\tilde G_1 \Lambda_1 F_m \Lambda_2 \tilde G^\dagger_2,
\end{equation}
where $\tilde G_1$ and $\tilde G_2$ are $N\times (N+L)$, and $\Lambda_{1,2}$ and
$F_m$ are $(N+L)\times(N+L)$ matrices with elements
\begin{equation}
  \tilde G_{1,lK}=G_l(a_K),\qquad  \tilde G_{2,lK}=G_l(b_K),
\end{equation}
\begin{equation}
  \Lambda_{1,IK}= \delta_{IK} e^{i\alpha_1 p_{K}+i E(p_{K})t}\qquad
  \Lambda_{2,IK}=\delta_{IK}  e^{-i\alpha_2 q_{K}-i E(q_{K})t},
\end{equation}
and finally 
\begin{equation*}
  F_{m,IK}= \frac{1-\left(\frac{b_I}{a_K}\right)^{N+L-m-1}}{1-\frac{b_I}{a_K}}.
\end{equation*}
The parametrization of the $a$ and $b$ variables reads
\begin{equation}
\begin{split}
a_I&=e^{ip_I},\qquad  p_I=I\frac{2\pi}{N+L}+ \frac{\pi +P_1}{N+L},\qquad
P_1=J_1\frac{2\pi}{L}\\
b_I&=e^{iq_I},\qquad  q_I=I\frac{2\pi}{N+L}+ \frac{\pi
  +P_2}{N+L},\qquad  
P_2=J_2\frac{2\pi}{L}.\\
\end{split}
\end{equation}
In those cases where $J_1\ne J_2$ we can use the Bethe equations to
write the matrix $F_m$ as
\begin{equation*}
  F_{m,IK}= \frac{1-
e^{i(P_2-P_1)}
\left(\frac{a_K}{b_I}\right)^{m+1}}{1-\frac{b_I}{a_K}}.
\end{equation*}
On the other hand, for $J_1=J_2$ we obtain
\begin{equation}
\label{Fdef}
  F_{m,IK}=(N+L)\delta_{IK}-\frac{a_K}{b_I}
\frac{1-\left(\frac{a_K}{b_I}\right)^{m+1}}{1-\frac{a_K}{b_I}}.
\end{equation}
Similar to the computation of the Loschmidt echo, we can insert the
matrices $V$ defined in \eqref{Vdef} into \eqref{efpT}. This gives an
alternative representation for the $O$-matrices:
\begin{equation}
\label{Odef}
O_m=H \Lambda_1 F_m \Lambda_2 H^\dagger,
\end{equation}
with $H$ given by \eqref{Hdef}. 

Formula \eqref{efpT} is exact and it computes the  EFP with
$\ordo(L^2A^2(N+L)^2N)\sim\ordo(N^7)$ steps. Although the power 
7 is quite large, this is still a huge simplification over the
original exponential sums. Moreover, depending on the situation, many
terms can be exactly identical which can lead to substantial improvements of the
numerical algorithm. 

If we restrict ourselves to calculate the space averaged EFP, it is enough the
keep terms with $J_1=J_2$:
\begin{equation}  
\label{avgEFP}
 \vev{\bar E_m(t)}\equiv 
\frac{1}{L}   \sum_{j=1}^L \vev{E_m(j,t)}=\frac{Z^2(N)}
{(AL(L+N)^{N-1})^2}
\sum_{J=1}^{L} 
\sum_{\alpha_1,\alpha_2=1}^{A}
 e^{i(\alpha_2-\alpha_1) P}  
\det O_m,
\end{equation}
where the matrix $O_m$ is defined
in \eqref{om} and  it still depends on $J$ and $\alpha_{1,2}$.

Further simplification is possible  when the initial
state is invariant under  translations by $p\ge 1$ sites. In these cases the sum in
\eqref{efpT} can be restricted to 
\begin{equation*}
  J_{1,2}=\frac{L}{p} \tilde J_{1,2},\qquad \tilde J_{1,2}=1,\dots,p.
\end{equation*}
In these cases the sum in \eqref{efpT} is more
manageable because its computational cost is $\ordo(N^5)$ instead of
the original $\ordo(N^7)$.
If full translational invariance holds ($p=1$) then only the
zero momentum sectors contribute:
\begin{equation}
\label{efpT2}
\vev{E_m(t)}=\frac{Z^2(N)}
{(AL(L+N)^{N-1})^2}
\sum_{\alpha_1,\alpha_2=1}^{A}
\det O_m.
\end{equation}

If the initial state is a pure Fock state, with partial or full
translational invariance, then depending on its actual structure the
number $A$ can be set to a constant value, and the cost of the resulting numerical
calculations becomes $\ordo(N^3)$. 
In appendix \ref{sec:equivalence} we also show that in these cases \eqref{efpT2} can
be transformed into a form which is identical to that obtained from
the mapping to the XX model introduced in the next section.

\section{Mapping to the XX chain}

\label{sec:XX}

The XX chain is a spin-$1/2$ model that can be mapped to free
fermions. It is defined through the Hamiltonian
\begin{equation}
  \label{HXX}
H_{\text{XX}}=-\sum_{j=1}^M( \sigma_j^+\sigma_{j+1}^-+\sigma_j^-\sigma_{j+1}^+-
h \sigma_j^z),
\end{equation}
where $\sigma_j^{+}$, $\sigma_j^{-}$, $\sigma_j^{z}$ are the Pauli matrices acting on site $j$.

In the following we describe two mappings between the phase
model and two different fermionic models .  The first one
connects the phase model with a slightly modified version of the XX
model; the modification consists of an extra non-local boundary
term. The second mapping connects the zero-momentum sectors of the
phase model and the original XX model. This mapping is exact, and
there are no extra boundary terms. In both cases we will work within
a subspace with fixed total particle number $N$. The length of the
bosonic and fermionic chains is $L$ and $M$, respectively.

As a first step we introduce the notations for the basis vectors.
In the same way as in \eqref{Fock} the 
states will be labeled by listing the
occupation numbers, but we add subscripts B (Bosons) and F (Fermions) to distinguish the
states of the two systems from each other:
\begin{equation*}
\begin{split}
  \ket{{\bf n}}_B=\ket{n_1}_{\text{B}}\otimes
  \ket{n_2}_{\text{B}}\otimes\cdots\otimes \ket{n_L}_{\text{B}},\qquad
  \sum_{j=1}^L n_j=N\\
  \ket{{\bf m}}_F=\ket{m_1}_{\text{F}}\otimes
  \ket{m_2}_{\text{F}}\otimes\cdots\otimes \ket{m_M}_{\text{F}},
\qquad
 \sum_{j=1}^M m_j=N.
\end{split}
\end{equation*}
In the fermionic case the allowed local Fock states are $\ket{0}_{\text{F}}$ and $\ket{1}_{\text{F}}$.

Now we define the mapping from the phase model to a modified XX
model. The main idea of this mapping first appeared in \cite{qboson-bog3}, but the
boundary conditions and the extra non-local boundary term were not
described there.

As a first step we introduce a mapping $P$ from a local bosonic space to an arbitrary product of
fermionic spaces
by the simple action
\begin{equation*}
  P \ket{n}_{\text{B}}=
\underbrace{\ket{1}_{\text{F}} \otimes \ket{1}_{\text{F}}\otimes\cdots\otimes \ket{1}_{\text{F}}}_{\text{n  times}}
\otimes \ket{0}_F.
\end{equation*}
In particular $P\ket{0}_{\text{B}}=\ket{0}_{\text{F}}$. 

The mapping can be extended to the full bosonic Hilbert space
naturally:
\begin{equation*}
P    \ket{{\bf n}}_B=
\otimes_{j=1}^L\ \big( P\ket{n_j}_{\text{B}}\big).
\end{equation*}
Examples are
\begin{equation*}
  P\ket{2,3,0,1}_B=\ket{1,1,0,1,1,1,0,0,1,0}_F\qquad P\ket{0,1,0}_B=\ket{0,1,0,0}_F
\end{equation*}

Another way to establish the mapping is through the coordinates of
the particles. If the bosonic particles have coordinates $x_j$, $j=1,\dots,N$, such
that $x_j\ge x_k$ for $j>k$, then the positions on the fermionic chain
are $y_j=x_j+j-1$. Note that this definition of the mapping is
suggested by the explicit wave function \eqref{coeff}, which becomes
a Slater determinant in the new coordinates.

Note that $P$ conserves the total particle number $N$ and it connects
the bosonic chain with length $L$ and the fermionic chain with length
$M=L+N$. However, the last site of the fermionic chain is not dynamic,
its state remains fixed to $\ket{0}_F$. Another way to formulate the
mapping is to delete the last site of the fermionic chain and then we
would get a bijection between the $N$-particle sector of the the phase
model of length $L$ and the $N$-particle sector of the fermionic chain
with length $N+L-1$. However, for future convenience we keep the last
site as well. 

Using the mapping $P$ we pull back the $q$-boson Hamiltonian to the
fermionic chain. We show that in the bulk we obtain the XX model, and
at the boundary a new term arises.
The fermionic Hamiltonian will be established by fixing its matrix
elements. In both the $q$-boson and XX models the Hamiltonian is a
sum of the hopping terms and the particle number term. The mapping
conserves number of particles, therefore the particle number operators
correspond to each other if the magnetic field in the XX model is
chosen to be $h=2$. On the other hand, the hopping terms
 are more complicated. 

A transition matrix element is non-zero in both models if and only if
there is a one-site hopping of one particle. On the bosonic side we
consider a hopping from site $j$ to site $j+1$. Let
\begin{equation*}
  \ket{a}_B=\ketB{\alpha}\otimes \ketB{a_j,a_{j+1}} \otimes \ketB{\beta}  \qquad\qquad
 \ket{a'}_B=
\ketB{\alpha}\otimes \ketB{a_j-1,a_{j+1}+1} \otimes \ketB{\beta}.
\end{equation*}
Here $\ketB{\alpha}$ (or $\ketB{\beta}$) is a state of the segment of
the chain from sites 1 to $j-1$ (or $j+2$ to $L$), respectively.
The Hamiltonian has a transition matrix element
\begin{equation*}
  _B\bra{a} H_{\text{B}} \ket{a'}_B =-1.
\end{equation*}
Then
\begin{equation*}
  \begin{split}
\ket{a}_F&=  P\ket{a}_B=\big(P \ketB{\alpha}\big)\otimes
\ketF{\underbrace{1,\dots,1}_{a_j-1\ \text{times} },1,0,\underbrace{1,\dots,1}_{a_{j+1}\ \text{times}  },0}
\otimes \big(P \ketB{\beta}\big)\\
\ket{a'}_F&=  P\ket{\tilde a}_B=\big(P \ketB{\alpha}\big)\otimes
\ketF{\underbrace{1,\dots,1}_{a_j-1\ \text{times} },0,1,\underbrace{1,\dots,1}_{a_{j+1}\ \text{times}  },0}
\otimes \big(P \ketB{\beta}\big).
  \end{split}
\end{equation*}
Therefore
\begin{equation*}
  _F\bra{a} H_{\text{XX}} \ket{a'}_F =-1.
\end{equation*}
Note that even though the matrix elements are the same, the real
location of the fermionic hopping depends on the particle
content of the states $\ketB{\alpha}$ and $\ketB{\beta}$.

Particle hoppings between the first and last sites of the $q$-boson
model need special treatment, and they result in new non-local terms
in the fermionic Hamiltonian. Let us investigate the two
vectors
\begin{equation*}
  \ket{b}_B=\ketB{b_1}\otimes \ketB{\gamma} \otimes \ketB{b_L}  \qquad\qquad
 \ket{b'}_B=\ketB{b_1-1}\otimes \ketB{\gamma} \otimes \ketB{b_L+1},
\end{equation*}
where $\ketB{\gamma}$ is an arbitrary vector of the chain segment between
sites 2 and $L-1$. We have the matrix element
\begin{equation*}
  _B\bra{b} H_{\text{B}} \ket{b'}_B =-1.
\end{equation*}
On the other hand, the mapping gives
\begin{equation*}
  \begin{split}
\ketF{b}=P \ketB{b}&=\ketF{\underbrace{1,\dots,1}_{b_1\ \text{times}},0}
\otimes \big(P\ketB{\gamma}\big)\otimes
\ketF{\underbrace{1,\dots,1}_{b_L\ \text{times}},0}\\
\ketF{b'}=P \ketB{b'}&=\ketF{\underbrace{1,\dots,1}_{b_1-1\ \text{times}},0}
\otimes \big(P\ketB{\gamma}\big)\otimes
\ketF{\underbrace{1,\dots,1}_{b_L+1\ \text{times}},0}.
  \end{split}
\end{equation*}
Alternatively these two vectors can be written as
\begin{equation*}
  \begin{split}
\ketF{b}=P \ketB{b}&=U\left[\ketF{\underbrace{1,\dots,1}_{b_1-1\ \text{times}},0}
\otimes \big(P\ketB{\gamma}\big)\otimes
\ketF{\underbrace{1,\dots,1}_{b_L\ \text{times}},0,1}\right]\\
\ketF{b'}=P \ketB{b'}&=\ketF{\underbrace{1,\dots,1}_{b_1-1\ \text{times}},0}
\otimes \big(P\ketB{\gamma}\big)\otimes
\ketF{\underbrace{1,\dots,1}_{b_L\ \text{times}},1,0},
  \end{split}
\end{equation*}
where $U$ is the periodic shift operator by one site to the
right. Therefore
\begin{equation}
  _{\text{F}}\bra{b}U H_{\text{XX}}\ketF{b'}=-1.
\end{equation}
and in the $N$-particle sector we have the similarity relation
between the Hamiltonians with sites $L$ and $M=N+L$ given by
\begin{equation}
  H_{\text{B}} =   P^{-1} H_{\text{XX}}^{mod} P
\end{equation}
with
\begin{equation}
  \label{eq:mapping}
H_{\text{XX}}^{mod}=  
-\left[\sum_{j=1}^{M-2} (  \sigma_j^+\sigma_{j+1}^-+\sigma_j^-\sigma_{j+1}^+)+
U \sigma_{M-1}^-\sigma_M^++
\sigma_{M-1}^+\sigma_M^- U^{-1}\right]+2N.
\end{equation}
Notice that $H_{\text{XX}}^{mod}$ respects the constraint that the
last site of the fermionic chain is fixed to $\ket{0}_F$.

Now we are in the position to establish a modified mapping $\tilde P$
which connects the zero-momentum sectors of the phase and XX
models. The idea is to symmetrize the original operator $P$ with
respect to translations, both on the bosonic and the fermionic
side. To define $\tilde P$ we first need to fix a basis in the
zero-momentum sectors. This can be achieved by selecting vectors
$\ket{\tilde a}_{B,F}$ in
the form of
\begin{equation}
\label{sumvec}
  \ket{\tilde a}_{B,F}=\frac{1}{\sqrt{p^a_{B,F}}} \left[
\ket{a}_{B,F}+U \ket{a}_{B,F}+U^2\ket{a}_{B,F}+\dots+ U^{p^a_{B,F}-1} 
\ket{a}_{B,F}\right],
\end{equation}
where $\ket{a}_{B,F}$ are Fock states of the bosonic and fermionic
chains that are invariant under translation by $p^a_{B,F}$
sites. Typically $p_B^a=L$ and
$p_F^a=N+L$, but we also need to treat the special cases with partial
translational invariance. An example is the zero-momentum N\'eel state
in the fermionic chain which has $p_F=2$ and is of the form
\begin{equation*}
  \ket{\tilde N}_F=\frac{1}{\sqrt{2}}\left[
\ket{1010\cdots}_F+\ket{0101\cdots}_F\right].
\end{equation*}
In both the bosonic and fermionic chains we can choose an arbitrary
member in the sum \eqref{sumvec} as the first (labelling) vector
$\ket{a}_{B,F}$. However, in the fermionic case we require that the last
site of $\ket{a}_{F}$ is in the state $\ket{0}_F$. This can always
be achieved by a finite number of translations because the fully occupied fermionic state does not occur in
the image of the original mapping $P$.

The mapping $\tilde P$ is constructed by associating 
\begin{equation*}
  \ket{\tilde a}_{F}=\tilde P \ket{\tilde a}_B
\end{equation*}
such that for the original vectors 
\begin{equation*}
  \ket{a}_F=P\ket{a}_B\quad  \text{ and }\quad p_F^a=\frac{N+L}{L}p_B^a.
\end{equation*}
The relation on the right above follows simply from the properties of
the mapping $P$. 
It is easy to see that $\tilde P$ establishes a bijection
between the zero-momentum basis vectors of the bosonic and fermionic
chains. Now we show that pulling back the $q$-boson Hamiltonian leads to
the XX model. 

First of all notice, that a one-particle hopping amplitude between two
vectors $\ket{a}_{B,F}$ and $\ket{b}_{B,F}$ is non-zero only in those cases when at least one of the
states has $p^{a,b}_B=L$ or $p^{a,b}_F=N+L$, respectively. If neither states are
periodic, then it can be seen
immediately  that the one-particle hopping amplitude is $-1$ both in
the bosonic and fermionic chains. Now we consider the cases when
$p^a_B<L$, $p^a_F<N+L$ and $p^a_B=L$, $p^a_F=N+L$. In these cases
breaking of the partial translational invariance is a result of a
single hopping somewhere along the chain. It is easy to see that
\begin{equation*}
  _B\langle{\tilde a}\left|H_B\right|{\tilde b}\rangle_B=-
  \frac{L}{\sqrt{p_B^ap_B^b}}=-\sqrt{\frac{L}{p_B^a}}\quad\text{and}\quad
 _F\langle{\tilde a}|H_{XX}|{\tilde b}\rangle_F=-
  \frac{N+L}{\sqrt{p_F^ap_F^b}}=-\sqrt{\frac{N+L}{p_F^a}}.
\end{equation*}
Using the relation $p^a_F=\frac{N+L}{L}p_B^a$, we obtain that the two matrix
elements are indeed the same. With this we have established that in
the zero-momentum sectors of each theory
\begin{equation}
  H_{\text{B}} =   \tilde P^{-1} H_{\text{XX}} \tilde P.
\end{equation}

It is useful to consider the physical $S$-matrix in the two models. In
the XX model we have $S=-1$, whereas in the phase model $S$ is given
by \eqref{S}. The apparent difference is easily explained by the
non-locality of the mapping $P$. Two particles that are on
neighboring sites of the phase model correspond to a string of $101$
on the XX chain. Therefore, exchanging positions of two particles in
the phase model is equivalent to performing a translation, an exchange
and another translation in the XX model. Multiplying the phase factors 
associated to these three steps, we obtain the bosonic $S$-matrix \eqref{S}.

\subsection{Other local operators}

The mappings between the models are highly non-local, therefore
there is no general correspondence between local operators. However,
we can find a limited set of operators that are local in both
models and whose space averages correspond to each other. Examples
include the emptiness formation probability (EFP) 
operators and the projection operators that measure the probability to
have a fixed number of particles on a given site.

First we consider the EFP. We define
local projection operators\footnote{The
  bosonic EFP operator has already been defined in Section
  \ref{sec:EFPt}. Here we redefine it with a subscript $B$ in order to have the same notations for the bosonic
  and fermionic operators.} $\Pi_{j,B}^0$ and $\Pi_{j,F}^0$ that project to
the subspaces where there is no particle on site $j$. Then the
position dependent $m$-site EFP is defined as
\begin{equation*}
  E_{m,B}(j)=\prod_{k=0}^{m-1} \Pi_{j+k,B}^0,\qquad\qquad
E_{m,F}(j)=\prod_{k=0}^{m-1} \Pi_{j+k,F}^0.
\end{equation*}
The average EFP is
\begin{equation*}
\bar  E_{m,B}=\frac{1}{L} \sum_{j=1}^L E_{m,B}(j),\qquad\qquad
\bar E_{m,F}=\frac{1}{M} \sum_{j=1}^M E_{m,F}(j).
\end{equation*}
It is easy to see that $P$ maps a local bosonic $m$-site EFP to a
local fermionic $m+1$ site EFP, but the position of the resulting fermionic
operator depends on the particle distribution of the bosonic state outside
the EFP. However, the averaged EFP's are mapped onto each other,
except for finite boundary contributions:
\begin{equation}
\label{EFPrel}
 \bar   E_{m,B}\approx \frac{M}{L}P^{-1}\bar
 E_{m+1,F}P=(1+n)P^{-1}\bar E_{m+1,F}P,
\end{equation}
where the factor $\frac{M}{L}=(1+n)$ arises from the change of the volume. Equality is
reached in the thermodynamic limit. 
Regarding the mapping $\tilde P$ connecting the zero-momentum sectors
it is easy to see that the relation
\begin{equation}
\label{EFPrel0}
  \bar   E_{m,B}=(1+n)\tilde P^{-1}\bar E_{m+1,F}\tilde P
\end{equation}
is exact both in finite volume and in the thermodynamic limit.

As a second example we consider the probability to have $m$ particles
on site $j$ of the bosonic model. To this end we introduce the general
projection operators $\Pi_{j,B}^m$ and $\Pi_{j,F}^m$, where
$m=0\dots\infty$ and $m=0,1$ in the bosonic and fermionic models,
respectively. We will also use the average of the bosonic operator
\begin{equation*}
\bar   \Pi_B^m=\frac{1}{L} \sum_{k=1}^L \Pi_{j,B}^m.
\end{equation*}
A bosonic site with $m$ particles is always mapped to a
continuous sequence of $m$ particles on the fermionic chain, preceded
and followed by an empty site:
\begin{equation*}
  P\ketB{\dots,m,\dots}=
\ketF{\dots,0,\underbrace{1,\dots,1}_{m\ \text{times}},0,\dots}.
\end{equation*}
We define the fermionic operators and their average
\begin{equation*}
  D_{m,F}(j)=\Pi_{j-1,F}^0\left( \prod_{k=0}^{m-1} \Pi_{j+k,F}^1 \right)
      \Pi_{j+m,F}^0,\qquad\qquad
\bar D_{m,F}=\frac{1}{M}\sum_{j=1}^M D_{m,F}(j).
\end{equation*}
Once again, the averaged operators correspond to each other except
for certain finite boundary terms:
\begin{equation}
\label{pol}
 \bar \Pi_B^m\approx (1+n) P^{-1}\bar D_{m,F} P.
\end{equation}
With regard to the mapping $\tilde P$ in the zero-momentum sector  we
have an exact relation
\begin{equation}
\label{pol0}
 \bar \Pi_B^m= (1+n) \tilde P^{-1}\bar D_{m,F} \tilde P.
\end{equation}
Note that \eqref{pol}-\eqref{pol0} give the EFP relations
\eqref{EFPrel}-\eqref{EFPrel0} at the special value $m=0$. 

\subsection{Formation probabilities in the XX chain}

As discussed in the previous subsection, the calculation of averaged formation probabilities
in the bosonic case can be directly recast in the language of the XX chain with Hamiltonian
\eqref{HXX}. The correspondence holds also for the time dependent quantities after a
quench and becomes especially useful in case of the translational invariant initial
states. 
Here we use this mapping to obtain simple determinant
formulas for the EFP, employing free-fermion techniques \cite{STN01,FA05}. On
the one hand, these will serve as a check of the bosonic results, but
they will also be used to obtain new results. For simplicity, we will
only consider cases when the bosonic state is translationally invariant.

The periodic XX chain can be transformed into a fermionic hopping model with Hamiltonian
\begin{equation}
H_F = -\sum_{j=1}^{M}(c^{\dag}_{j+1}c_j + c^{\dag}_jc_{j+1}-h(2c^\dag_jc_j-1)),
\label{HF}
\end{equation}
where, in the sector of even particle number $N=\sum_j c^\dag_jc_j$,
the boundary condition is anti-periodic in the fermion operators, $c_{M+1}=-c_1$.
Note that the lengths $M$ and $L$ of the fermionic and bosonic chains are related as
$M=L+N$. Since the term proportional to $h$ in \eqref{HF} is just a constant,
it drops out from the time evolution and one can set $h=0$. The Hamiltonian
becomes diagonal after a Fourier transform
\begin{equation}
H_F = -\sum_{k=1}^{M} 2\cos (q_k) c^{\dag}_{q_k}c_{q_k}, \qquad
q_{k} = \frac{\pi}{M}\left(2k-1\right),
\end{equation}
where the allowed values of momenta are set by the anti-periodic boundary condition.

Starting from an initial Fock state, all the information about the time evolved state
is encoded in the fermionic correlation matrix elements $C_{mn}(t) = \langle c_m^\dag(t) c_n(t)\rangle$.
Using the diagonal form of $H_F$, the time evolution of the fermionic operators reads
\begin{equation}
c_n(t)=\sum_l U_{nl}(t) c_l \, , \quad 
U_{nl}(t)=\frac 1 M \sum_k e^{-iq_k(n-l)} e^{i2t \cos q_k}.
\label{cnt}
\end{equation}
From here on, we will work directly in the thermodynamic limit $M\to\infty$, where
the matrix elements of the unitary time evolution operator are given via Bessel functions
as $U_{nl}(t)=i^{n-l}J_{n-l}(2t)$. The time dependent correlation matrix elements then follow as
\cite{EP07}
\begin{equation}
C_{mn}(t)=i^{n-m} \sum_{k,l} i^{k-l} J_{m-k}(2t) J_{n-l}(2t) C_{kl}(0) \, .
\label{ct}
\end{equation}

The knowledge of the above two-point functions allows us to calculate the expectation
value of arbitrary products of creation and annihilation operators through Wick's theorem.
In particular, the position dependent $m$-site EFP is given by \cite{STN01,FA05}
\begin{equation}
\vev{E_{m,F}(j,t)} = \langle \prod_{k=0}^{m-1} c^{\phantom\dag}_{j+k}(t)c^{\dag}_{j+k}(t)\rangle =
\det (\identity - C_{I_{j,m}}(t)),
\label{efpf}
\end{equation}
where $I_{j,m} = \left[ j, j+m-1\right]$ and for any interval $I$ the
matrix $C_{I}$ is the reduced correlation matrix with
both its rows and columns restricted to $I$. Finally, the averaged
bosonic EFP is obtained as
\begin{equation}
\vev{\bar E_{m,B}(j,t)} =
\frac{1}{L}\sum_{j=1}^{N+L}
\det (\identity - C_{I_{j,m}}(t)).
\label{efpf2}
\end{equation}
If the bosonic initial state is such that its fermionic picture is
invariant under translations under $\nu$ sites then we have the
simpler formula
\begin{equation}
  \vev{\bar E_{m,B}(j,t)} =
\frac{1+n}{\nu}\sum_{j=1}^{\nu}
\det (\identity - C_{I_{j,m}}(t)).
\label{efpf3}
\end{equation}
For the sake of completeness, we show in Appendix \ref{sec:equivalence} that \eqref{efpf3} can be
derived directly from formula \eqref{efpT2} without invoking the mapping to
the XX model.

One can also obtain determinant formulas for the various
fermionic string formation probabilities $\vev{D_{m,F}(j,t)}$ that are related via
\eqref{pol} to the probability of finding $m$ bosons at a given site. In particular, for
$m=1$ one has
\begin{equation}
\vev{D_{1,F}(j,t)} = \det(\identity-C_{I_1}) - \det(\identity-C_{I_2}),
\label{d1fjt}
\end{equation}
where $I_1=\left[j-1\right] \cup \left[ j+1\right]$ and $I_2=\left[j-1,j+1\right]$.
Note, that this is just the EFP on two next-nearest-neighbor sites $j-1$ and $j+1$
minus the EFP on all three consecutive sites, which indeed gives the probability
of finding a $010$ string.

Finally, we consider the string-probability $D_{2,F}(j,t)$ on sites $\left[j-1,j+2\right]$
in the time evolved state. This can again be written in terms of EFP's as
\begin{equation}
\vev{D_{2,F}(j,t)} = \det(\identity-C_{I_1}) - \det(\identity-C_{I_2})
- \det(\identity-C_{I_3})  + \det(\identity-C_{I_4})
\label{d2fjt}
\end{equation}
on the corresponding subsets
\begin{align*}
I_1&=\left[j-1\right] \cup \left[j+2\right], &
I_2&=\left[j-1,j\right] \cup \left[j+2\right], \\
I_3&=\left[j-1\right] \cup \left[j+1,j+2\right], &
I_4&=\left[j-1,j+2\right].
\end{align*}
Formulas \eqref{d1fjt}-\eqref{d2fjt} will be used in the next section to derive the
bosonic occupation probabilities in a specific quench situation.

\section{Translationally invariant initial states}

\label{sec:global}

In this section we calculate exact time evolution starting from 
translationally invariant initial states.
We deal with homogeneous initial states: 
\begin{equation*}
  \ket{\psi_n}_B\equiv\ket{n,n,n,n,\cdots}_B
\end{equation*}
On the one hand these are 
natural candidates, because they could be prepared in experiments or
other numerical simulations. On the other hand they are also
convenient for our purposes because their overlaps have the form
\eqref{talanilyen} so that the calculations of the previous sections
can be applied.

Our methods enable us to investigate both finite size effects and the
infinite volume limit. Concerning finite size effects, we can address
the question of particle propagation on the background induced by the
quench. In a finite volume global quench we expect that
 the physical observables will be very close to their infinite volume
limit until a time $t\approx L/2v^*$, where $v^*$ is interpreted as the velocity of
 propagation through the highly excited stationary state reached after
 the quench. The physical interpretation can be given using the
 semi-classical reasoning of \cite{cardy-calabrese-quench1}: as a
 result of the quench quasi-particles are emitted at each site, and
 finite size effects arise only at times when two particles have
 traveled around the volume and meet at the opposite side. Our
 numerical data enables us to give estimates for the velocities
 $v^*$. We stress however, that these can not be considered as
 rigorous answers, because there are no sharp wave fronts associated
 with the finite size effects.

Concerning the infinite volume and infinite time limit, we can check
the predictions of the GGE. In \cite{sajat-qboson} it was argued that
in the $q$-boson model the GGE should give a correct description of the stationary state
after the quench\footnote{It might be more accurate to use the term
  ``predictions of the Diagonal Ensemble (DE)'', because an actual
  Generalized Gibbs Ensemble is not constructed. However, for
  historical reasons and to conform with the terminology of \cite{JS-CGGE},
  we continue to use the expression ``predictions of the GGE''.}. The
main argument was that in the $q$-boson  model the
initial values of the conserved charges completely determine what
types of states can populate the system after the quench, in
particular there is a one-to-one correspondence between the charges
and the Bethe root densities. After dephasing, all local operators
approach the averages in these Bethe states, and the averages only depend on the
root densities. Therefore, the charges in the initial state completely
determine the long-time limit of local observables. 
This was
explicitly  checked in
\cite{sajat-qboson} for the particular quench that will be discussed in \ref{sec:1111}. In the present work we
can go further and consider more general initial states. 

In  \cite{sajat-qboson} it was shown that the mean
values of the EFP operators in Bethe states  can be 
expressed using the charges, for example 
\begin{equation*}
  \bra{\Omega}E_1 \ket{\Omega}=\frac{1}{1+n}\left(1-| \bra{\Omega}Q_1 \ket{\Omega}|^2\right),
\end{equation*}
where $\ket{\Omega}$ is an arbitrary on-shell Bethe state.
It was also shown that in pure Fock states
the expectation values of the charges are all zero except for the
particle number operator. Therefore it is very easy to obtain the GGE
predictions for the limit of the EFP, 
for example
\begin{equation}
\label{EFP1pred}
 \lim_{t\to\infty} \vev{E_1(t)}=\frac{1}{1+n},
\end{equation}
which should be valid for any quench starting from an initial state which is
a pure Fock state. 
 In particular, we will verify this for the states
$|\psi_n \rangle_B$ at the end of this section.

\subsection{Initial state $\ket{\psi_1}_B=\ket{1,1,1,1\dots}_B$}

\label{sec:1111}

This is probably the simplest initial state with particle density
$n=N/L=1$: there is exactly one particle at each site. 
The initial state corresponds to the particular component of the Bethe
vector \eqref{coeff} with $x_j=j$, therefore the overlap can be
written as
\begin{equation}
\label{ov11}
\skalarszorzat{\Psi_0}{\{a\}_N}  =
\det_N\Big( (a_k)^{2j}\Big)=\prod_j a_j^2 \prod_{j<k}  (a_k^2-a_j^2)=
 \prod_{j<k}  (a_k^2-a_j^2),
\end{equation}
where we used that the overlap is non-zero for zero-momentum states
only.
This simple structure of the overlap made possible the direct
computation of the time-evolution of the 1-EFP
in \cite{sajat-qboson}. The following result was obtained in finite volume:
\begin{equation}
\label{jolettez0}
\vev{E_1(t)}=
\frac{1}{2}-
\frac{1}{2}\left(\frac{1}{N}\sum_{j}   
\cos(4\cos(p_j)t)\right)-\frac{1}{2}\left|\frac{1}{N}\sum_{j}
\sin(4\cos(p_j)t)e^{ip_j}\right|^2,
\end{equation}
where
\begin{equation}
\label{pj}
  p_j=\frac{\pi (2j-1)}{2L}.
\end{equation}
Taking the infinite volume limit with fixed $t$ leads to
\begin{equation}
\label{jolettez}
\begin{split}
\vev{E_1(t)}=
\frac{1}{2}-\frac{1}{2}
\left(\int_{0}^\pi \frac{dp}{\pi} \cos(4\cos(p)t)\right)^2
-\frac{1}{2}
\left|\int_{0}^\pi \frac{dp}{\pi} \sin(4\cos(p)t)e^{ip}\right|^2.
\end{split}
\end{equation}
The GGE prediction \eqref{EFP1pred} gives
\begin{equation*}
   \lim_{t\to\infty} \vev{E_1(x,t)}=\frac{1}{2},
\end{equation*}
and this is clearly confirmed by \eqref{jolettez}, as it was already
observed in \cite{sajat-qboson}.

Quite interestingly the symmetrized mapping to the XX model gives
\begin{equation}
\label{NAN}
  \tilde P\ketB{1,1,\dots,1}=\frac{1}{\sqrt{2}}\left[
\ketF{1,0,1,0,\cdots}+\ketF{0,1,0,1,\cdots}
\right].
\end{equation}
Therefore, this particular situation corresponds to the N\'eel-to-XX quench studied extensively in 
\cite{gritsev-stb-neel-to-xx-first,gritsev-demler-stb-quench-osszf-neel-to-xx,andraschko-sirker,XX-quench-brockmann},
given that the initial state is chosen to be the translationally
invariant combination of the N\'eel and Anti-N\'eel states. 
These papers concentrated on the time evolution of the staggered
anti-ferromagnetic order, the spin-spin correlations, and the
Loschmidt echo.

Below we also show how to compute
\eqref{jolettez0} in the fermionic language
introduced in section \ref{sec:XX}. However, before turning to the XX
model we first consider the Loschmidt amplitude. 
In appendix \ref{sec:Lo1} we compute  it using the
general formalism introduced in section \ref{sec:exact}. Our finite
volume result for the amplitude reads
\begin{equation}
\label{lecho11a}
      \lecho=\left[
 \prod_{j=1}^N  \cos(2\cos(p_j)t)+
 \prod_{j=1}^N  \sin(2\cos(p_j)t)
\right], 
\end{equation}
where $p_j$ is defined in \eqref{pj}. This agrees with the
corresponding result of
\cite{XX-quench-brockmann}.
The second term in \eqref{lecho11a} is not present in the N\'eel to XX
quench, because it corresponds to the finite volume N\'eel to
Anti-N\'eel transition. 
If we fix $t$ then
in the $L\to\infty$ limit only the first term survives. 
The physical reason for this is that in infinite volume the time
evolution generated by the Hamiltonian 
can not shift the whole chain by one site in finite time. Therefore, in the
TDL limit at finite $t$ we have
\begin{equation}
  \lim_{L\to\infty} \frac{1}{L}\log  |\lecho|=
\frac{1}{\pi}\int_0^{\pi}  \log|  \cos(2\cos(p)t)|  dp.
\end{equation}
We note that even though the finite volume amplitude is exactly equal
to that computed in \cite{andraschko-sirker}, there is a factor of 2
in the Loschmidt echo per site. This factor arises from the re-scaling
of the volume through the mapping between the two models.

Now we turn to the evaluation of the EFP and the local occupation
probabilities.  Formula \eqref{jolettez0} could be derived easily from  \eqref{efpT2} too,
for example by using \eqref{Odef} for a simple evaluation
of the determinants. However, it is instructive to re-derive it 
in the fermionic language
introduced in section \ref{sec:XX}.

We are interested in the 2-site EFP $\vev{E_{2,F}(j,t)}$, which can be obtained via the
time-dependent fermionic correlation matrix \eqref{ct} using the determinant formula \eqref{efpf}.
The correlation matrix at time $t=0$ is diagonal with alternating entries
$C_{2k-1,2k-1}(0)=1$ and $C_{2k,2k}(0)=0$. Substituting into \eqref{ct} and using
the Bessel function addition theorems \cite{Abramowitz}
\begin{equation}
\sum_{k=-\infty}^{\infty} J_{k+m}(z)J_{k+n}(z) = \delta_{m,n}, \qquad
\sum_{k=-\infty}^{\infty} (-1)^k J_{k+m}(z)J_{k+n}(z) = (-1)^m J_{n-m}(2z),
\label{add1}
\end{equation}
one obtains the following simple expression for the time evolved matrix elements
\begin{equation}
C_{m,n}(t) = \frac{i^{n-m}}{2} \left[ \delta_{m,n} - (-1)^m J_{n-m}(4t)\right].
\label{ct1}
\end{equation}
Although the matrix elements are invariant only to 2-site translations, it is easy to
see that the $2 \times 2$ determinants defining the 2-site EFP in \eqref{efpf}
are completely translationally invariant. Therefore, one has for the average EFP
\begin{equation}
\vev{\bar E_{2,F}(t)} = \frac{1}{4} (1-J^2_0(4t)-J^2_1(4t)).
\label{efpf1}
\end{equation}
Comparing to the bosonic result in \eqref{jolettez}, one obtains the relation
\eqref{EFPrel} with the volume re-scaling factor $1+n=2$. Moreover, as
discussed in the previous section,
one has an exact match even with the finite volume result \eqref{jolettez0},
which is easily obtained from \eqref{efpf1} by replacing the Bessel
functions in \eqref{efpf1} with
the finite-$M$ propagators in \eqref{cnt} and setting $M=2L$.

\begin{figure}[!ht]
  \centering
 \includegraphics[scale=0.7]{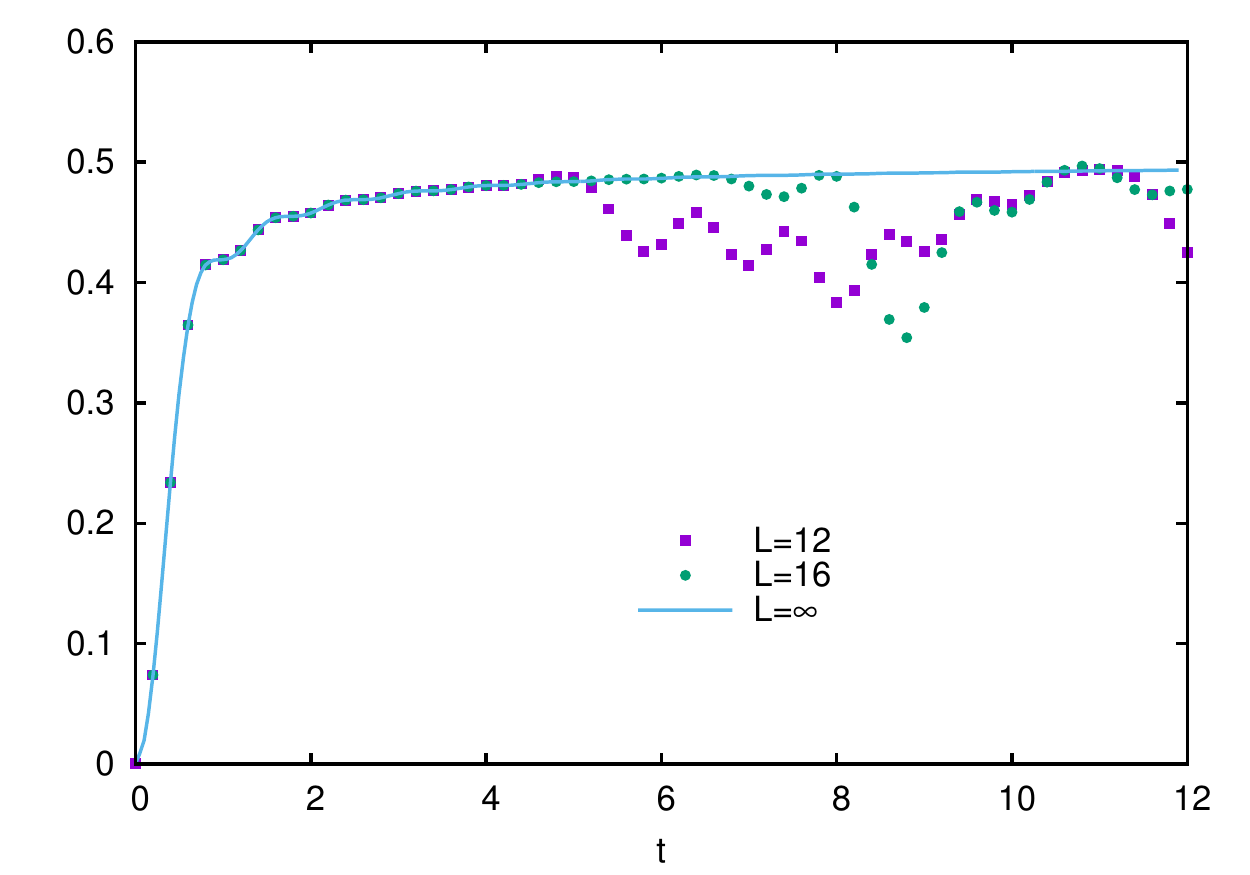} 
\caption{Time evolution of the 1-site EFP in the phase model for the
  initial state $\ket{\psi_1}_B=\ket{1,1,1,1,\cdots}_B$.}
\label{FigGlobalF1A}
\end{figure}

In Fig. \ref{FigGlobalF1A} we plotted the numerical data for the
bosonic 1-EFP. We can see that the finite volume time
evolution  follows the infinite volume curve until a time
$t^*=L/2v^*$ which
grows linearly with the volume. The speed $v^*$ of the propagation could
be estimated from the difference between the finite and infinite
volume results, or simply by looking at the numerical data for large
volumes. 
We choose the latter method and read off the approximate value
$v^* \approx 1$. Note that the maximum speed of
particles in vacuum was $v_{\text{max}}=2$, which is the group
velocity as calculated from the energy-momentum relation
\eqref{enmom}. This slowing down with a factor of 2 can be explained
by the properties of the mappings $P$ and $\tilde P$ between the
bosonic and fermionic models. In the XX chain information always
propagates with a maximal speed of $v=2$, regardless of the local spin
content. This follows from the fact that the model is effectively
free. On the other hand, the mappings are non-local and they
involve a re-scaling of the volume whose extent depends on the local
particle density. In fact, a segment of the bosonic chain with length
$l$ and density $n$ is mapped to a segment of the XX model with length
$l(1+n)$. Therefore, information in the bosonic chain can not
propagate faster than  $v^*(n)=2/(1+n)$. In the present case this leads
to $v^*=1$, in agreement with the numerical data.

We continue by calculating the expectation value of the average
string-probability $\vev{\bar D_{1,F}(t)}$
which, in the bosonic chain, corresponds to the probability of having
a single boson at a site. We stress that this is a new result,
  which could not have been obtained using the formalism of Section
  \ref{sec:exact}, because the matrix elements of the corresponding
  operators are not known.
The fermionic $010$ string probability is position dependent, however, due to 2-site translational invariance,
it is enough to consider $j=1$ and $j=2$. 
Using the matrix elements \eqref{ct1} and expanding the determinants in \eqref{d1fjt},
one finds for the average string probability
\begin{equation}
\vev{\bar D_{1,F}(t)}= 
\frac{1}{2} \sum_{j=1,2} \vev{D_{1,F}(j,t)} =
\frac{1}{8}(1+3J_0^2+2J_1^2-J_2^2),
\end{equation}
where for brevity the arguments $(4t)$ of the Bessel functions were suppressed.
For the bosonic occupation probability we thus obtain
\begin{equation}
\label{d010}
\vev{\Pi^1_{B}(t)}= 
\frac{1}{4}(1+3J_0^2+2J_1^2-J_2^2).
\end{equation}
Similarly, one can also calculate the $0110$-string probability $\vev{D_{2,F}(j,t)}$,
which is, in fact, translational invariant due to the symmetry of the problem
and thus one has to consider $j=1$ only. Evaluating the determinants in \eqref{d2fjt}
is straightforward but rather tedious. After re-scaling one arrives at the lengthy formula
\begin{align}
\label{d0110}
\vev{\Pi^2_{B}(t)} =\frac{1}{8} \left[ \right.
&(1-J^2_0)(1-J^2_0-3J^2_1 - J^2_3) - 2(1+J^2_0)J^2_2 + 4J_0 J_1 J_2 (J_1-J_3)
\nonumber \\
&\left. +2J_1J_3(J^2_1-J^2_2) + J^2_1J^2_3 + (J^2_1 + J^2_2)^2 + 4(J^2_1+J^2_2) \right].
\end{align}

\begin{figure}[!ht]
  \center
 \includegraphics[scale=0.7]{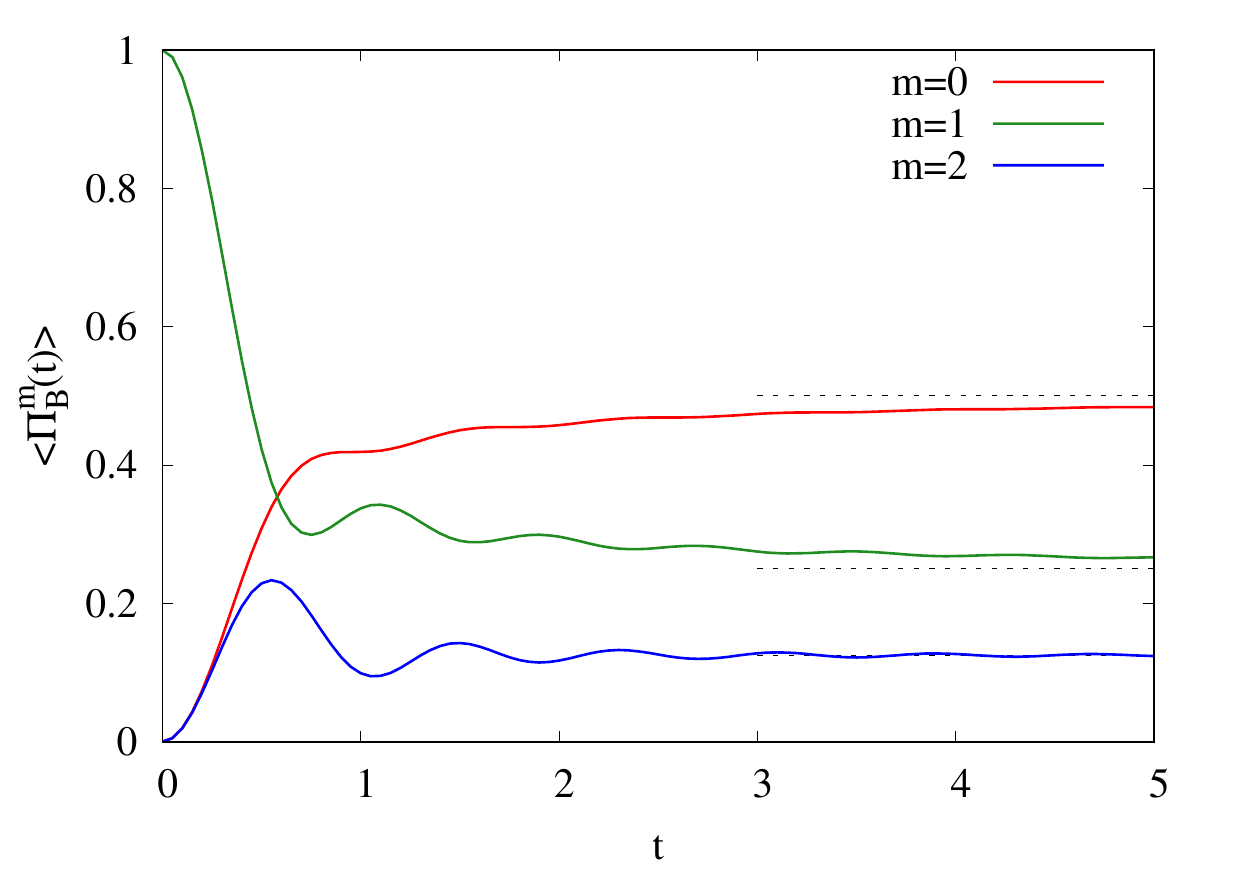} 
\caption{Time evolution of the 
  probability to have $m$ bosons on a single site for
  the initial state $\ket{\psi_1}_B=\ket{1,1,1,1,\cdots}_B$.
The bosonic result is obtained from the averaged fermionic string
probabilities $\vev{\bar D_{m,F}(t)}$ for $m=0,1,2$
multiplied by the factor $1+n=2$,
see Eq. \eqref{pol0}.
The horizontal dashed lines correspond to the $t\to\infty$ limit.
}
\label{fig:pstring}
\end{figure}

The bosonic occupation probabilities $\vev{\Pi^m_B(t)}$ are shown in
Fig. \ref{fig:pstring} for $m=0,1,2$. 
One observes that the curves
relax rapidly to their stationary values, given by $1/2, 1/4, 1/8$, respectively. This suggests
that the number of bosons has a geometric distribution in the stationary state. In the fermionic
picture it is easy to verify that this is indeed the case. In fact, the reduced density matrix $\rho_I$
of an arbitrary finite interval $I$ in the stationary state is given via the reduced correlation matrix
$C_I$ as \cite{PE09}
\begin{equation}
\rho_I \propto \exp(\sum_{i,j \in I} H_{i,j} c^{\dag}_{i}c_j), \qquad
H = \ln \frac{1-C_I}{C_I},
\end{equation}
where
$C_{I} = \lim_{t\to\infty} C_I(t)$. From Eq. \eqref{ct1} one has immediately $C_I = \identity/2$
and hence the local steady state is a Gibbs state at infinite temperature.
In turn, this implies that all string configurations have equal probabilities and thus
\begin{equation*}
\lim_{t\to\infty}\vev{\bar D_{m,F}(t)} = 2^{-(m+2)}.
\end{equation*}
We stress that equations \eqref{d010}-\eqref{d0110} are new results
of this work.

%%%%%%%%%%%%%%%%%%%%%%%%%%%
%%%%%%%%%%%%%%%%%%%%%%%%%%%

\subsection{Initial state $\ket{\psi_2}_B=\ket{2,2,2,2\dots}_B$}

As a second example, we consider the initial state with exactly 2 particles at each site. 
The overlaps
are of the form \eqref{talanilyen} with 
\begin{equation*}
  G_j(a_k)=a_k^{j+x_j},\ \text{with}\quad
  x_{j}=\left[\frac{j+1}{2}\right],
\end{equation*}
and $[x]$ denotes the integer part. 
In this case the time evolution of the EFP can be computed either from
formula \eqref{efpT2} or by the mapping to the XX chain. In the
following we present the latter computation.

The symmetrized mapping gives
\begin{equation}
\tilde P\ketB{2,2,\dots}=\frac{1}{\sqrt{3}}\left[
\ketF{1,1,0,1,1,0,\cdots}+
\ketF{0,1,1,0,1,1,\cdots}+
\ketF{1,0,1,1,0,1,\cdots}\right]
\end{equation}
For simplicity we will work in the thermodynamic limit.
Since the initial density has now a 3-site periodicity, we need a more general addition
theorem of Bessel functions \cite{Abramowitz}
\begin{equation}
\begin{split}
\sum_{k=-\infty}^{\infty} J_{n+k}(z) J_{k}(z) \cos (k \alpha)
&=J_{n}(2 \sin (\alpha/2) z) \cos (n (\pi-\alpha)/2) 
\\
\sum_{k=-\infty}^{\infty} J_{n+k}(z) J_{k}(z) \sin (k \alpha)
=&J_{n}(2 \sin (\alpha/2) z) \sin (n  (\pi-\alpha)/2).
\label{add2}
\end{split}
\end{equation}
Indeed, setting $\alpha=2\pi/3$, we obtain two equations involving the
various 3-periodic sums of products of Bessel functions. Supplementing the set
of equations with the completeness relation on the l.h.s of Eq. \eqref{add1},
the system can be solved and the result can be used to evaluate the matrix elements in
\eqref{ct}. In particular, the diagonal elements are obtained by choosing
$n=0$ in \eqref{add2} with the result
\begin{equation}
C_{m,m}(t) = 
\begin{cases}
\frac{1}{3}(2+ J_{0}(2\sqrt{3}t))& \mbox{if $m=3l+1$ or $m=3l+2$} \\
\frac{2}{3}(1- J_{0}(2\sqrt{3}t))& \mbox{if $m=3l$}.
\end{cases}.
\end{equation}
Similarly, the elements in the first off-diagonal follow from solving the set
of equations for $n=1$ and read
\begin{equation}
C_{m,m+1}(t) = 
\begin{cases}
0 & \mbox{if $m=3l+1$} \\
iJ_{1}(2\sqrt{3}t)/\sqrt{3} & \mbox{if $m=3l+2$} \\
-iJ_{1}(2\sqrt{3}t)/\sqrt{3} & \mbox{if $m=3l$}.
\end{cases}
\end{equation}
Evaluating the determinants and taking the average, one arrives at
\begin{equation}
\vev{\bar E_{2,F}(t)}=\frac{1}{3}\sum_{j=1}^{3} \vev{E_{2,F}(j,t)}=
\frac{1}{9} \left[1-J^2_0(2\sqrt{3}t) -2J^2_1(2\sqrt{3}t) \right].
\end{equation}
For the bosonic 1-EFP we thus obtain
\begin{equation}
\label{efp1B22}
\vev{E_{1,B}(t)}=
\frac{1}{3} \left[1-J^2_0(2\sqrt{3}t) -2J^2_1(2\sqrt{3}t) \right].
\end{equation}
The finite volume result has the same form, with the Bessel functions
replaced by the finite volume propagators \eqref{cnt}.

\begin{figure}[!ht]
  \centering
\includegraphics[scale=0.7]{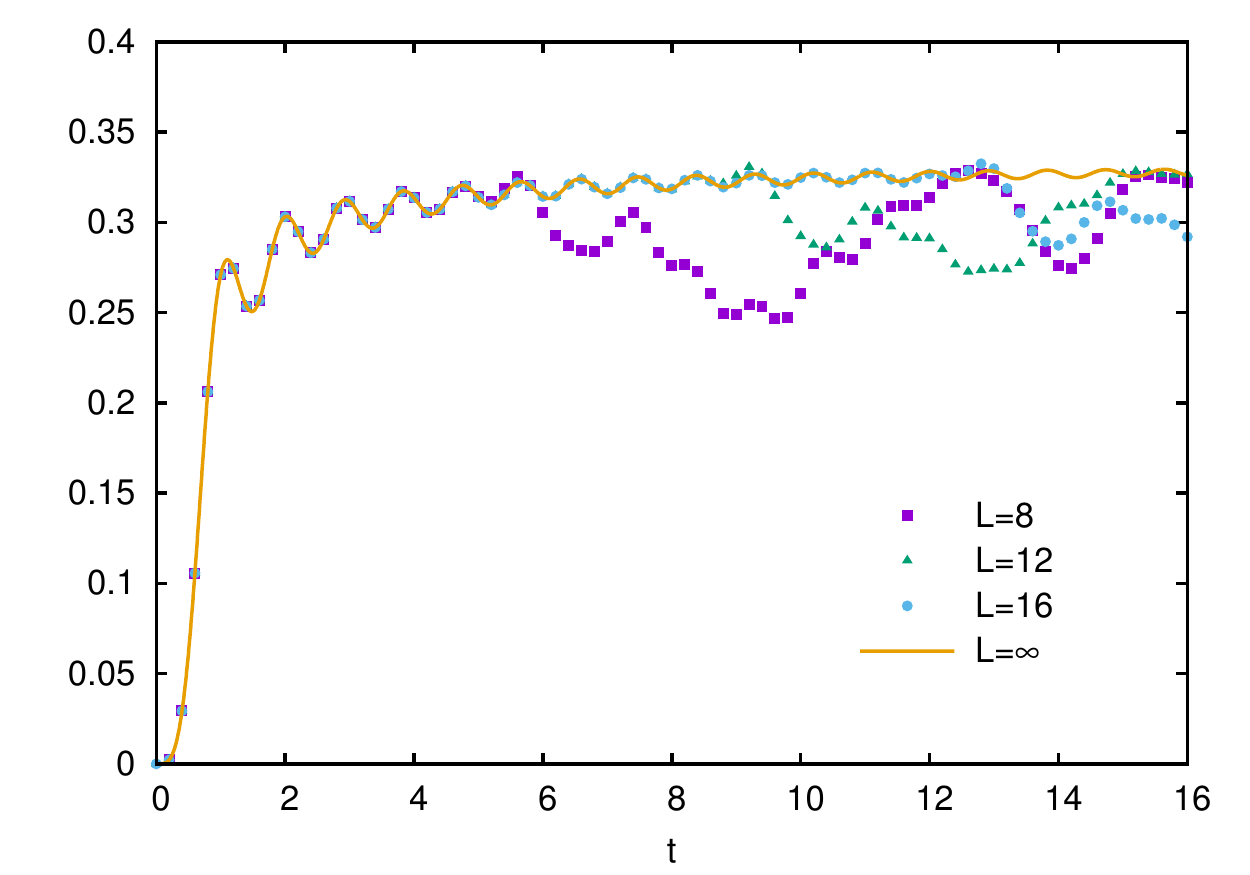}
\caption{Time evolution of the 1-site EFP in the phase model for the
  initial state $\ket{\psi_2}_B=\ket{2,2,2,2,\cdots}_B$.}
\label{FigGlobalF2A}
\end{figure}

In Fig. \ref{FigGlobalF2A} we plot the numerical data for the 1-EFP in
finite and infinite volume. 
Concerning the maximal speed of wave propagation,  
we can read off the approximate value $v^*\approx 2/3$. Similar to our
previous example, the slowing down of the wave propagation is
consistent with the re-scaling of the volume caused by $\tilde P$ with a factor of 3. 

Note that the long-time limit of the 1-EFP in \eqref{efp1B22}  is given by $1/3$
which agrees with the GGE prediction \eqref{EFP1pred}. In fact, the prediction can
even be verified for initial states $|\psi_n\rangle_B$ with arbitrary $n$.
As observed already for $n=1$ and $n=2$, it is not difficult to prove in general that
for $t\to\infty$ the off-diagonal matrix elements of the fermionic correlation
matrix vanish, while the diagonal elements give the average fermionic density
\begin{equation}
\lim_{t \to \infty} C_{m,m}(t) = \frac{n}{n+1}, \qquad
\lim_{t \to \infty} C_{m,m+1}(t) =0.
\end{equation}
Substituting into \eqref{efpf3}, one immediately finds
\begin{equation}
\lim_{t \to \infty} \vev{E_{1,B}(t)} = \frac{1}{1+n}.
\end{equation}

In Appendix \ref{sec:Lo2} we also computed the time evolution of the
Loschmidt amplitude, both in finite and infinite volume. The formulas
are lengthy and we refrain from repeating them here. However, we note
that the intermediate result \eqref{Lalakul3} leads to a relatively
simple calculation even in this case. In fact, it seems that 
regarding the Loschmidt amplitude the
bosonic formulation always leads to simpler derivations 
than the fermionic methods.

\section{Conclusions}

In this work we investigated time evolution of physical observables in
the phase model, which is the  $q\to\infty$
limit of the $q$-boson model. We set up a framework to calculate the finite volume
Loschmidt 
echo and the emptiness formation probability in non-equilibrium
situations. One of our main results is equation \eqref{efpT}, which
computes the exact EFP in polynomial ($\ordo(N^7)$) time, which is a tremendous
simplification as compared to the exponential sum over all states in
the Hilbert space. Analogously, the Loschmidt echo can be obtained
with $\ordo(N^5)$ steps. 
The result \eqref{efpT}  applies whenever the overlaps with the initial state take the
form \eqref{talanilyen}. Pure Fock states in the local coordinate
basis satisfy this requirement. Such states are natural candidates to study
because they could be prepared in experiments or other independent
numerical simulations. 

The fact that the EFP could be obtained with a cost of $\ordo(N^7)$
steps shows that the complexity of the phase model is
between that of a free theory and a generic Bethe Ansatz solvable
model. In technical terms, this unique situation arises because the
scattering phase shift \eqref{S} can be factorized. This leads to a
special type of Bethe equation \eqref{BeQ}, which implies that the
one-particle rapidities can be chosen freely from an enlarged set of
$N+L$ solutions, as long as they obey an
overall constraint coming from the total momentum
quantization. Unfortunately, this simplification only emerges in the
$q\to\infty$ limit, which restricts the scope of our method.

We also considered a non-local mapping between the phase model and the
XX chain. In its simplest form, the mapping connects the zero-momentum
sectors of the two theories, such that the Hamiltonians and space
averages of certain other local operators are mapped onto each
other. In situations that are not translationally invariant the
mapping can not be used to give information about the space dependence
of the observables, it applies only to the averaged
operators. Therefore it is important to stress that the phase model
should not be considered as ``equivalent to a free theory''.

In Section \ref{sec:global} we considered examples where the initial
state is translationally invariant. We derived new results for the
Loschmidt amplitude, the EFP, and local bosonic occupation
probabilities using both the bosonic methods of section \ref{sec:exact}
and the mapping to the XX chain. In these formulas both the
thermodynamic limit and the long time limit could be performed quite easily;
in those cases where a GGE prediction
was available for the stationary values in the TDL, the
prediction was confirmed. 
We stress that, even though our
results pertain to very special situations, they are among the very
few that exactly solve time evolution and rigorously prove the
validity of the GGE predictions in an interacting many-body
system.

In a future publication we plan to apply our result \eqref{efpT} in
situations that are not translationally invariant. Examples include
global quenches starting from initial states that have periodic
structure, for example $\ketB{\Psi_0}=\ketB{1,0,1,0,\cdots}$. Formula
\eqref{efpT} can be readily applied in those cases, giving the
numerically exact finite volume EFP. However, it is a
challenging open problem to obtain the thermodynamic limit of the
space dependent EFP in such cases. 

Another class of problems to be
investigated is that of joining two subsystems with different initial
states. This can be achieved for example by choosing initial states of
the form 
$\ketB{\Psi_0}=\ketB{1,1,\cdots,1,2,2,\cdots,2}$. Our formulas deal
with periodic boundary conditions, therefore this situation
corresponds to having two initial domain walls on a circle. Time
evolution will induce a particle and energy current at
the two boundaries, and we plan to investigate the properties of these
currents and the wave-front propagation into the bulk of the two subsystems. 

\vspace{1cm}
{\bf Acknowledgments} 

\bigskip

The authors are grateful to Zolt\'an Zimbor\'as for suggesting to study
these problems, to G\'abor Tak\'acs for helping to construct the
mapping $\tilde P$ introduced in Section \ref{sec:XX}, to Michael
Brockmann for discussions about the relation to the XX model and in
particular for drawing our attention to Fig. 4 of \cite{qboson-bog3}
which describes the mapping $P$, and finally to M\'arton Kormos for
very useful discussions which inspired us to construct the finite
volume techniques of Section \ref{sec:exact}.

V.E. acknowledges funding from the Austrian Science Fund (FWF) through
Lise Meitner Project No. M1854-N36. 

\appendix

\section{Analytic results for the Loschmidt amplitude}

Here we compute exact analytic results for the Loschmidt amplitude
using the formula \eqref{Lalakul3}. For simplicity we subtract an
irrelevant constant from the Hamiltonian and will work with the
one-particle dispersion relation
\begin{equation*}
  E(p)=-2\cos(p).
\end{equation*}
This only affects the phase of the Loschmidt amplitude.

\subsection{Initial state $\ket{\psi_1}_B=\ket{1,1,1,1,\dots}_B$}

\label{sec:Lo1}

We have $N=L$, the state is translationally invariant and only the $J=0$
sector contributes. The initial coordinates of the particles are $x_j=j$. 
Then the matrix $H$ is $N\times 2N$ and 
\begin{equation}
  H_{jK}=\frac{1}{\sqrt{N}}\sum_{m} e^{-iq_j m}  e^{i2 p_K m}.
\end{equation}
We will use the definitions
\begin{equation}
  q_j=\frac{2\pi}{N} j -\frac{\pi}{N},\quad j=1\dots N
  \quad\text{and}\quad p_K=\frac{\pi}{N}K-\frac{\pi}{2N},\quad
  K=1\dots 2N.
\end{equation}
This gives
\begin{equation}
  H_{jK}=\sqrt{N}(\delta_{j,K}+\delta_{j+N,K}).
\end{equation}
This way the matrix product in \eqref{Lalakul3} gives
\begin{equation*}
  (H \Lambda H^\dagger )_{jk}=
N\delta_{j,k}\left(\Lambda_{j,j}+\Lambda_{j+N,j+N}\right)=
N\delta_{jk} \left(e^{-iE(q_j/2)t}e^{i\alpha q_j/2}+e^{-iE(q_j/2+\pi)t}e^{i\alpha (q_j/2+\pi)}\right).
\end{equation*}
We have
\begin{equation*}
 \det  (H \Lambda H^\dagger )=N^N
  \begin{cases}
   \prod_{j=1}^N 2 \cos(2\cos(q_j/2)t) e^{i\alpha q_j/2} & 
 \text{for  even } \alpha \\
   \prod_{j=1}^N 2i \sin(2\cos(q_j/2)t) e^{i\alpha q_j/2} & 
 \text{for odd } \alpha
  \end{cases}.
\end{equation*}
For the sum of the momenta we obtain
\begin{equation}
  \sum_{j=1}^N q_j=-\pi+\frac{2\pi}{N}  \frac{N(N+1)}{2}=\pi N.
\end{equation}
Therefore
\begin{equation}
  \prod_{j=1}^N e^{i\alpha q_j/2} = e^{i\alpha \pi N/2}=
  \begin{cases}
    1 & \text{for  even } \alpha \\ 
(-1)^{N/2} &  \text{for odd } \alpha
  \end{cases}.
\end{equation}
This gives
\begin{equation*}
 \det  (H \Lambda H^\dagger )=2^NN^N
  \begin{cases}
   \prod_{j=1}^N  \cos(2\cos(q_j/2)t)  & 
 \text{for  even } \alpha\\
  \prod_{j=1}^N  \sin(2\cos(q_j/2)t)  & 
 \text{for odd } \alpha 
  \end{cases}.
\end{equation*}
All quantities are periodic in $\alpha$ with period 2, so we can
set $A=2$ in \eqref{Lalakul3} leading to
\begin{equation}
\label{lecho11}
      \lecho=\left[
 \prod_{j=1}^N  \cos(2\cos(q_j/2)t)+
 \prod_{j=1}^N  \sin(2\cos(q_j/2)t)
\right].
\end{equation}

\subsection{Initial state $\ket{\psi_2}_B=\ket{2,2,2,2,\dots}_B$}

\label{sec:Lo2}

In this case 
we have $N=2L$, only the $P=0$ sector contributes and
 \begin{equation*}
   x_j=
   \begin{cases}
(j+1)/2 & \text{for odd } j \\
     j/2 & \text{for even } j 
   \end{cases}.
 \end{equation*}
We have the definitions
\begin{equation}
\begin{split}
  q_j=\frac{2\pi}{2L} j -\frac{\pi}{2L},\quad j=1\dots 2L\quad \text{and}\quad 
p_K=\frac{2\pi}{3L}K-\frac{\pi}{3L},\quad K=1\dots 3L.
\end{split}
\end{equation}
The matrix $H$ is $2L\times 3L$ and its elements are
\begin{equation}
  H_{jK}=\frac{1}{\sqrt{N}}
(1-e^{i(q_j-p_K)})
\sum_{m=1}^{L} 
e^{-iq_j 2m} e^{ip_K 3m}=\frac{\sqrt{N}}{2}
(1-e^{i(q_j-p_K)})
(\delta_{j,K} \text{ mod } L).
\end{equation}
For the product in \eqref{Lalakul3} we obtain a block-diagonal matrix,
whose determinant is
\begin{equation}
  \det (H \Lambda H^\dagger )=\left(\frac{L}{2}\right)^N
\prod_{j=1}^L \det M^j,
\end{equation}
where each $M^j$ is a $2\times 2$ matrix with elements given by
\begin{equation}
\begin{split}
  M^j_{ab}&=
\sum_{c=0}^2  (1-e^{i(q_{j+aL}-p_{j+cL})})  (1-e^{-i(q_{j+bL}-p_{j+cL})})
\Lambda_{j+cL,j+cL} 
%\\
%&=\sum_{c=0}^2  (1-e^{i(q_{j+aL}-p_{j+cL})})(1-e^{-i(q_{j+bL}-p_{j+cL})})
%e^{i\alpha p_{j+cL}}e^{-i E(p_{j+cL})t}.
\end{split}
\end{equation}
For the sub-determinants we get
\begin{equation*}
\begin{split}
\det M^j&=
\sum_{c,d=0}^2 16 \sin\left(\frac{q_j-p_{j+cL}}{2}\right)\cos\left(\frac{q_j-p_{j+dL}}{2}\right)
\sin\left(\frac{p_{j+dL}-p_{j+cL}}{2}\right)    \Lambda_{j+cL}\Lambda_{j+dL}\\
&=
\sum_{c,d=0}^2 16 \sin\left(\frac{q_j-p_{j+cL}}{2}\right)\cos\left(\frac{q_j-p_{j+dL}}{2}\right)
\sin\left(\frac{(c-d)\pi}{3}\right)    \Lambda_{j+cL}\Lambda_{j+dL}\\
&=
\sum_{c,d=0}^2 16 \sin\left(\frac{q_j}{6}-\frac{c\pi}{3}\right)
\cos\left(\frac{q_j}{6}-\frac{d\pi}{3}\right)
\sin\left(\frac{(c-d)\pi}{3}\right)
e^{2i\alpha p_{j}}e^{i\alpha (c+d) 2\pi/3 }e^{-i (E(p_{j+cL})+E(p_{j+dL}))t}.
\end{split}
\end{equation*}
All quantities are  $\alpha$-periodic with a period of 3, therefore we
can set $A=3$: 
\begin{equation*}
       \lecho=
\sum_{\alpha=1}^3 
e^{i\alpha  L 2\pi/3} 
\prod_{j=1}^L z_\alpha(q_j),
\end{equation*}
where
\begin{equation*}
\begin{split}
  z_\alpha(q_j)=&\frac{4}{9}
\sum_{c,d=0}^2 \left[ \sin\left(\frac{q_j}{6}-\frac{c\pi}{3}\right)
\cos\left(\frac{q_j}{6}-\frac{d\pi}{3}\right)
\sin\left(\frac{(c-d)\pi}{3}\right)\times 
\right.\\
&\hspace{3cm}\times \left.
e^{i\alpha \frac{ (c+d) 2\pi}{3} }
e^{2it \left(\cos\left(\frac{2q_{j}+c2\pi}{3}\right)+
\cos\left(\frac{2q_{j}+d2\pi}{3}\right)\right)}\right].
\end{split}
\end{equation*}
In the thermodynamic limit only the term with $\alpha=3$ survives and
this leads to 
\begin{equation}
  \lim_{L\to\infty} \frac{1}{L}\log  |\lecho|=
\frac{1}{\pi}\int_0^{\pi}dq\  \log|  z(q) |,
\end{equation}
where
\begin{equation*}
\begin{split}
  z(q)=&\frac{4}{9}
\sum_{c,d=0}^2 \left[ \sin\left(\frac{q}{6}-\frac{c\pi}{3}\right)
\cos\left(\frac{q}{6}-\frac{d\pi}{3}\right)
\sin\left(\frac{(c-d)\pi}{3}\right)
e^{2it \left(\cos\left(\frac{2q+c2\pi}{3}\right)+
\cos\left(\frac{2q+d2\pi}{3}\right)\right)}\right].
\end{split}
\end{equation*}

\section{Direct equivalence of the bosonic and fermionic calculations of the EFP}

\label{sec:equivalence}

Here we prove that the fermionic results for the EFP can be obtained
directly from the bosonic formula \eqref{efpT}. We require that the
initial state is translationally invariant. 
In order to conform with the notations in the main text we will use
the convention that the lower and upper case indices take values $1,\dots,N$
and $1,\dots,(N+L)$, respectively. 

We introduce the $(N+L)\times (N+L)$ Fourier-transform matrix as
\begin{equation*}
  U_{IJ}=\frac{1}{\sqrt{N+L}}e^{ip_JI},
\end{equation*}
where the $p_J$ variables are defined in \eqref{asol2}. It is easy to
see that $U$ is unitary. Note that this matrix is different from $V$
introduced in \eqref{Vdef}, which performs a Fourier transform over $N$ points.

Our goal is to calculate the EFP from expression \eqref{efpT2}. In the
present calculation  it is useful to treat the two factors in
the $\Lambda$ matrices
separately. Inserting $1=UU^\dagger=U^\dagger U$ 
we write
with some abuse of notation 
\begin{equation*}
O_m=
\tilde G \Lambda_1 F_m \Lambda_2 \tilde G^\dagger=
\tilde Ge^{i\alpha_1p} U^\dagger U e^{iE(p)t}  F_m e^{-iE(p)t} U^\dagger U  e^{-i\alpha_2p} \tilde G ^\dagger.
\end{equation*}
It is easy to see that
\begin{equation*}
(\tilde Ge^{i\alpha_{1,2}p}  U^\dagger)_{kJ}=\sqrt{N+L}\delta_{J,k+x_k+\alpha_{1,2}}.
\end{equation*}
Using \eqref{Fdef} we write 
\begin{equation*}
\left(e^{iE(p)t}  F_m e^{-iE(p)t} \right)_{JK}=(
N+L)\delta_{JK}-\sum_{\beta=1}^{m+1}
\frac{e^{iE(p_J)t}}{e^{iE(p_K)t}} \frac{e^{i\beta p_J}}{e^{i \beta p_K}}.
\end{equation*}
Multiplying with $U$ and $U^\dagger$ we get
\begin{equation*}
\left(Ue^{iE(p)t}  F_m e^{-iE(p)t} U^\dagger\right)_{JK}=(
N+L)\delta_{JK}-(N+L)
\sum_{\beta=1}^{m+1} U_{J+\beta}^*(t)U_{K+\beta}(t),
\end{equation*}
where 
\begin{equation*}
  U_\beta(t)=\frac{1}{{N+L}} \sum_{J=1}^{N+L} e^{-i\beta p_J} e^{-iE(p_J)t},
\end{equation*}
which is equal to the fermionic propagator introduced in \eqref{cnt},
up to an irrelevant constant shift in the energy eigenvalue.

Finally
\begin{equation*}
  O_{m,jk}=(N+L)^2\left[\delta_{jk}\delta_{\alpha_1,\alpha_2}-\sum_{\beta=1}^{m+1}
U_{j+x_j+\alpha_1+\beta}^*(t)U_{k+x_k+\alpha_2+\beta}(t)\right].
\end{equation*}
If $\alpha_1\ne \alpha_2$ then the determinant vanishes except in the very
small chain with $N=m+1$. We don't consider this case and set
$\alpha_1=\alpha_2$. 

The determinant of $O_m$ can be computed using the following general
identity, which is valid for arbitrary $\kappa\le N$:
\begin{equation}
\label{jokistetel}
  \det_N (\delta_{jk}+\sum_{\beta=1}^\kappa a^{(\beta)}_jb^{(\beta)}_k)_{j,k=1..N}
=\det_{\kappa}  (\delta_{\beta,\gamma}+\sum_{j=1}^N a^{(\beta)}_jb^{(\gamma)}_j)_{\beta,\gamma=1..\kappa}.
\end{equation}
In the present case we get
\begin{equation*}
 \det O_m=(N+L)^{2N} \det_{m+1} K, 
\end{equation*}
where
\begin{equation*}
  K_{\beta\gamma}=\delta_{\beta\gamma}-
\sum_{j=1}^N U^*_{j+x_j+\alpha+\beta}(t)U_{j+x_j+\alpha+\gamma}(t).
\end{equation*}
If the coordinates $y_j=j+x_j$ form a set of periodicity $\nu$, then we can set
$A=\nu$ and obtain
\begin{equation}
\label{trinvEFP}
  \vev{E_m(t)}=\frac{(1+n)^2}{\nu^2}
\sum_{\alpha=1}^{\nu}
\det_{m+1} K.
\end{equation}
If the initial state is $\ket{\psi_n}_B$ then $\nu=1+n$ 
and the sum in \eqref{trinvEFP} describes the averaging over the $1+n$
components of the fermionic state $\ket{\psi_n}_F=\tilde P
\ket{\psi_n}_B$. It is easy to check that \eqref{trinvEFP} exactly agrees with formula \eqref{efpf3},
which was derived using free fermion techniques.

\addcontentsline{toc}{section}{References}
\bibliography{../../../pozsi-general}

\providecommand{\href}[2]{#2}\begingroup\raggedright\begin{thebibliography}{10}

\bibitem{Silva-quench-colloquium}
A.~Polkovnikov, K.~Sengupta, A.~Silva, and M.~Vengalattore,
  ``\textit{Colloquium} : Nonequilibrium dynamics of closed interacting quantum
  systems,'' \href{http://dx.doi.org/10.1103/RevModPhys.83.863}{{\em Rev. Mod.
  Phys.} {\bf 83} (2011)  863--883}, \href{http://arxiv.org/abs/1007.5331}{{\tt
  arXiv:1007.5331 [cond-mat.stat-mech]}}.

\bibitem{QM-out-of-equilibrium-review}
J.~{Eisert}, M.~{Friesdorf}, and C.~{Gogolin}, ``{Quantum many-body systems out
  of equilibrium},'' \href{http://dx.doi.org/10.1038/nphys3215}{{\em Nature
  Physics} {\bf 11} (2015)  124--130},
  \href{http://arxiv.org/abs/1408.5148}{{\tt arXiv:1408.5148 [quant-ph]}}.

\bibitem{rigol-gge}
M.~Rigol, V.~Dunjko, V.~Yurovsky, and M.~Olshanii, ``Relaxation in a Completely
  Integrable Many-Body Quantum System: An Ab Initio Study of the Dynamics of
  the Highly Excited States of 1D Lattice Hard-Core Bosons,''
  \href{http://dx.doi.org/10.1103/PhysRevLett.98.050405}{{\em Physical Review
  Letters} {\bf 98} (2007) no.~5, 050405},
  \href{http://arxiv.org/abs/arXiv:cond-mat/0604476}{{\tt
  arXiv:cond-mat/0604476}}.

\bibitem{vidal-itebd1}
G.~{Vidal}, ``{Efficient Simulation of One-Dimensional Quantum Many-Body
  Systems},'' \href{http://dx.doi.org/10.1103/PhysRevLett.93.040502}{{\em
  Physical Review Letters} {\bf 93} (2004) no.~4, 040502},
  \href{http://arxiv.org/abs/quant-ph/0310089}{{\tt quant-ph/0310089}}.

\bibitem{vidal-itebd2}
J.~{Jordan}, R.~{Or{\'u}s}, G.~{Vidal}, F.~{Verstraete}, and J.~I. {Cirac},
  ``{Classical Simulation of Infinite-Size Quantum Lattice Systems in Two
  Spatial Dimensions},''
  \href{http://dx.doi.org/10.1103/PhysRevLett.101.250602}{{\em Physical Review
  Letters} {\bf 101} (2008) no.~25, 250602},
  \href{http://arxiv.org/abs/cond-mat/0703788}{{\tt cond-mat/0703788}}.

\bibitem{sajat-oTBA}
B.~{Pozsgay}, M.~{Mesty{\'a}n}, M.~A. {Werner}, M.~{Kormos}, G.~{Zar{\'a}nd},
  and G.~{Tak{\'a}cs}, ``{Correlations after Quantum Quenches in the XXZ Spin
  Chain: Failure of the Generalized Gibbs Ensemble},''
  \href{http://dx.doi.org/10.1103/PhysRevLett.113.117203}{{\em Physical Review
  Letters} {\bf 113} (2014) no.~11, 117203},
  \href{http://arxiv.org/abs/1405.2843}{{\tt arXiv:1405.2843
  [cond-mat.stat-mech]}}.

\bibitem{quench-action}
J.-S. {Caux} and F.~H.~L. {Essler}, ``{Time Evolution of Local Observables
  After Quenching to an Integrable Model},''
  \href{http://dx.doi.org/10.1103/PhysRevLett.110.257203}{{\em Physical Review
  Letters} {\bf 110} (2013) no.~25, 257203},
  \href{http://arxiv.org/abs/1301.3806}{{\tt arXiv:1301.3806
  [cond-mat.stat-mech]}}.

\bibitem{caux-stb-LL-BEC-quench}
J.~De~Nardis, B.~Wouters, M.~Brockmann, and J.-S. Caux, ``Solution for an
  interaction quench in the Lieb-Liniger Bose gas,''
  \href{http://dx.doi.org/10.1103/PhysRevA.89.033601}{{\em Physical Review A}
  {\bf 89} (2014)  033601}, \href{http://arxiv.org/abs/1308.4310}{{\tt
  arXiv:1308.4310 [cond-mat.stat-mech]}}.

\bibitem{JS-oTBA}
B.~{Wouters}, J.~{De Nardis}, M.~{Brockmann}, D.~{Fioretto}, M.~{Rigol}, and
  J.-S. {Caux}, ``{Quenching the Anisotropic Heisenberg Chain: Exact Solution
  and Generalized Gibbs Ensemble Predictions},''
  \href{http://dx.doi.org/10.1103/PhysRevLett.113.117202}{{\em Physical Review
  Letters} {\bf 113} (2014) no.~11, 117202},
  \href{http://arxiv.org/abs/1405.0172}{{\tt arXiv:1405.0172
  [cond-mat.str-el]}}.

\bibitem{JS-CGGE}
E.~{Ilievski}, J.~{De Nardis}, B.~{Wouters}, J.-S. {Caux}, F.~H.~L. {Essler},
  and T.~{Prosen}, ``{Complete Generalized Gibbs Ensembles in an Interacting
  Theory},'' \href{http://dx.doi.org/10.1103/PhysRevLett.115.157201}{{\em
  Physical Review Letters} {\bf 115} (2015) no.~15, 157201},
  \href{http://arxiv.org/abs/1507.02993}{{\tt arXiv:1507.02993 [quant-ph]}}.

\bibitem{finite-qa}
V.~{Alba} and P.~{Calabrese}, ``{The quench action approach in finite
  integrable spin chains},'' {\em ArXiv e-prints} (2015)  ,
  \href{http://arxiv.org/abs/1512.02213}{{\tt arXiv:1512.02213
  [cond-mat.str-el]}}.

\bibitem{ABACUS}
J.~{Caux}, ``{Correlation functions of integrable models: A description of the
  ABACUS algorithm},'' \href{http://dx.doi.org/10.1063/1.3216474}{{\em Journal
  of Mathematical Physics} {\bf 50} (2009) no.~9, 095214},
  \href{http://arxiv.org/abs/0908.1660}{{\tt arXiv:0908.1660
  [cond-mat.str-el]}}.

\bibitem{JS-Milos}
M.~{Panfil} and J.-S. {Caux}, ``{Finite-temperature correlations in the
  Lieb-Liniger one-dimensional Bose gas},''
  \href{http://dx.doi.org/10.1103/PhysRevA.89.033605}{{\em Physical Review A}
  {\bf 89} (2014) no.~3, 033605}, \href{http://arxiv.org/abs/1308.2887}{{\tt
  arXiv:1308.2887 [cond-mat.quant-gas]}}.

\bibitem{JS-asympt-from-QA}
J.~{De Nardis}, L.~{Piroli}, and J.-S. {Caux}, ``{Relaxation dynamics of local
  observables in integrable systems},''
  \href{http://dx.doi.org/10.1088/1751-8113/48/43/43FT01}{{\em Journal of
  Physics A Mathematical General} {\bf 48} (2015)  43FT01},
  \href{http://arxiv.org/abs/1505.03080}{{\tt arXiv:1505.03080
  [cond-mat.quant-gas]}}.

\bibitem{korepin-Book}
V.~Korepin, N.~Bogoliubov, and A.~Izergin, {\em Quantum inverse scattering
  method and correlation functions}.
\newblock Cambridge University Press, 1993.

\bibitem{karol-hab}
K.~K. {Kozlowski}, ``{Asymptotic analysis and quantum integrable models},''
  {\em ArXiv e-prints} (2015)  , \href{http://arxiv.org/abs/1508.06085}{{\tt
  arXiv:1508.06085 [math-ph]}}.

\bibitem{kluemper-goehmann-finiteT-review}
J.~{Sato}, B.~{Aufgebauer}, H.~{Boos}, F.~{G{\"o}hmann}, A.~{Kl{\"u}mper},
  M.~{Takahashi}, and C.~{Trippe}, ``{Computation of Static Heisenberg-Chain
  Correlators: Control over Length and Temperature Dependence},''
  \href{http://dx.doi.org/10.1103/PhysRevLett.106.257201}{{\em Physical Review
  Letters} {\bf 106} (2011) no.~25, 257201},
  \href{http://arxiv.org/abs/1105.4447}{{\tt arXiv:1105.4447
  [cond-mat.str-el]}}.

\bibitem{js-hosszu-kvencs}
M.~{Brockmann}, B.~{Wouters}, D.~{Fioretto}, J.~{De Nardis}, R.~{Vlijm}, and
  J.-S. {Caux}, ``{Quench action approach for releasing the N\'{e}el state into
  the spin-1/2 XXZ chain},''
  \href{http://dx.doi.org/10.1088/1742-5468/2014/12/P12009}{{\em Journal of
  Statistical Mechanics: Theory and Experiment} {\bf 12} (2014)  12009},
  \href{http://arxiv.org/abs/1408.5075}{{\tt arXiv:1408.5075
  [cond-mat.str-el]}}.

\bibitem{sajat-QA-GGE-hosszu-cikk}
M.~{Mesty{\'a}n}, B.~{Pozsgay}, G.~{Tak{\'a}cs}, and M.~A. {Werner},
  ``{Quenching the XXZ spin chain: quench action approach versus generalized
  Gibbs ensemble},''
  \href{http://dx.doi.org/10.1088/1742-5468/2015/04/P04001}{{\em Journal of
  Statistical Mechanics: Theory and Experiment} {\bf 4} (2015)  1},
  \href{http://arxiv.org/abs/1412.4787}{{\tt arXiv:1412.4787
  [cond-mat.stat-mech]}}.

\bibitem{jacopo-michael-hirota}
E.~{Ilievski}, E.~{Quinn}, J.~{De Nardis}, and M.~{Brockmann}, ``{String-charge
  duality in integrable lattice models},'' {\em ArXiv e-prints} (2015)  ,
  \href{http://arxiv.org/abs/1512.04454}{{\tt arXiv:1512.04454
  [cond-mat.stat-mech]}}.

\bibitem{qboson-izergin-kitanine-bog}
N.~Bogoliubov, A.~Izergin, and N.~Kitanine, ``Correlation functions for a
  strongly correlated boson system,''
  \href{http://dx.doi.org/http://dx.doi.org/10.1016/S0550-3213(98)00038-8}{{\em
  Nuclear Physics B} {\bf 516} (1998) no.~3, 501 -- 528}.

\bibitem{qboson-bog1}
N.~M. {Bogoliubov}, ``{Boxed plane partitions as an exactly solvable boson
  model},'' \href{http://dx.doi.org/10.1088/0305-4470/38/43/002}{{\em Journal
  of Physics A Mathematical General} {\bf 38} (2005)  9415--9430},
  \href{http://arxiv.org/abs/cond-mat/0503748}{{\tt cond-mat/0503748}}.

\bibitem{qboson-bog2}
N.~M. {Bogoliubov}, ``{Enumeration of plane partitions and the algebraic Bethe
  anzatz},'' \href{http://dx.doi.org/10.1007/s11232-007-0012-5}{{\em
  Theoretical and Mathematical Physics} {\bf 150} (2007)  165--174}.

\bibitem{qboson-bog3}
N.~Bogoliubov and J.~Timonen, ``Correlation functions for a strongly coupled
  boson system and plane partitions,''
  \href{http://dx.doi.org/10.1098/rsta.2010.0322}{{\em Philosophical
  Transactions of the Royal Society of London A: Mathematical, Physical and
  Engineering Sciences} {\bf 369} (2011) no.~1939, 1319--1333}.

\bibitem{KIEU94}
V.~E. Korepin, A.~G. Izergin, F.~H.~L. Essler, and D.~B. Uglov, ``Correlation
  function of the spin-1/2 XXX antiferromagnet,''
  \href{http://dx.doi.org/10.1016/0375-9601(94)90074-4}{{\em Phys. Lett. A}
  {\bf 190} (1994)  182}, \href{http://arxiv.org/abs/cond-mat/9403066}{{\tt
  cond-mat/9403066}}.

\bibitem{EFIK95}
F.~H.~L. Essler, H.~Frahm, A.~G. Izergin, and V.~E. Korepin, ``Determinant
  representation for correlation functions of spin-1/2 XXX and XXZ Heisenberg
  magnets,'' \href{http://dx.doi.org/10.1007/BF02099470}{{\em Comm. Math.
  Phys.} {\bf 174} (1995)  191},
  \href{http://arxiv.org/abs/hep-th/9406133}{{\tt hep-th/9406133}}.

\bibitem{sajat-qboson}
B.~{Pozsgay}, ``{Quantum quenches and generalized Gibbs ensemble in a Bethe
  Ansatz solvable lattice model of interacting bosons},''
  \href{http://dx.doi.org/10.1088/1742-5468/2014/10/P10045}{{\em Journal of
  Statistical Mechanics: Theory and Experiment} {\bf 10} (2014)  45},
  \href{http://arxiv.org/abs/1407.8344}{{\tt arXiv:1407.8344
  [cond-mat.stat-mech]}}.

\bibitem{qbozon-bog-bullough1}
N.~M. Bogoliubov and R.~K. Bullough, ``A q-deformed completely integrable Bose
  gas model,'' \href{http://dx.doi.org/10.1088/0305-4470/25/14/020}{{\em
  Journal of Physics A: Mathematical and General} {\bf 25} (1992) no.~14,
  4057}.

\bibitem{qbozon-bog-bullough2}
N.~M. {Bogoliubov} and R.~K. {Bullough}, ``{Completely integrable model of
  interacting q-bosons},''
  \href{http://dx.doi.org/10.1016/0375-9601(92)91129-F}{{\em Physics Letters A}
  {\bf 168} (1992)  264--269}.

\bibitem{qbozon-bog-bullough3}
N.~M. Bogoliubov, R.~K. Bullough, and G.~D. Pang, ``Exact solution of a
  \textit{q}-boson hopping model,''
  \href{http://dx.doi.org/10.1103/PhysRevB.47.11495}{{\em Phys. Rev. B} {\bf
  47} (1993)  11495--11498}.

\bibitem{qboson-keiichi}
K.~{Shigechi} and M.~{Uchiyama}, ``{Boxed skew plane partition and integrable
  phase model},'' \href{http://dx.doi.org/10.1088/0305-4470/38/48/003}{{\em
  Journal of Physics A Mathematical General} {\bf 38} (2005)  10287--10306},
  \href{http://arxiv.org/abs/cond-mat/0508090}{{\tt cond-mat/0508090}}.

\bibitem{AK02}
A.~G. Abanov and V.~E. Korepin, ``On the probability of ferromagnetic strings
  in antiferromagnetic spin chains,''
  \href{http://dx.doi.org/10.1016/S0550-3213(02)00899-4}{{\em Nucl. Phys. B.}
  {\bf 647} (2002)  565}, \href{http://arxiv.org/abs/cond-mat/0206353}{{\tt
  cond-mat/0206353}}.

\bibitem{KMST02}
N.~Kitanine, J.~M. Maillet, N.~Slavnov, and V.~Terras, ``Emptiness formation
  probability of the XXZ spin-1/2 Heisenberg chain at $\Delta = 1/2$,''
  \href{http://dx.doi.org/10.1088/0305-4470/35/27/102}{{\em J. Phys. A: Math.
  Gen.} {\bf 35} (2002)  L385}, \href{http://arxiv.org/abs/hep-th/0201134}{{\tt
  hep-th/0201134}}.

\bibitem{KMST02b}
N.~Kitanine, J.~M. Maillet, N.~Slavnov, and V.~Terras, ``Large distance
  asymptotic behaviour of the emptiness formation probability of the XXZ
  spin-1/2 Heisenberg chain,''
  \href{http://dx.doi.org/10.1088/0305-4470/35/49/102}{{\em J. Phys. A: Math.
  Gen.} {\bf 35} (2002)  L753}, \href{http://arxiv.org/abs/hep-th/0210019}{{\tt
  hep-th/0210019}}.

\bibitem{KLNS03}
V.~E. Korepin, S.~Lukyanov, Y.~Nishiyama, and M.~Shiroishi, ``Asymptotic
  behavior of the emptiness formation probability in the critical phase of XXZ
  spin chain,'' \href{http://dx.doi.org/10.1016/S0375-9601(03)00616-9}{{\em
  Phys. Lett. A} {\bf 312} (2003)  21},
  \href{http://arxiv.org/abs/cond-mat/0210140}{{\tt cond-mat/0210140}}.

\bibitem{Kozlowski08}
K.~K. Kozlowski, ``On the emptiness formation probability of the open XXZ
  spin-1/2 chain,''
  \href{http://dx.doi.org/10.1088/1742-5468/2008/02/P02006}{{\em J. Stat.
  Mech.} (2008)  P02006}, \href{http://arxiv.org/abs/0708.0433}{{\tt
  0708.0433}}.

\bibitem{Cantini12}
L.~Cantini, ``Finite size emptiness formation probability of the XXZ spin chain
  at $\Delta=-1/2$,''
  \href{http://dx.doi.org/10.1088/1751-8113/45/13/135207}{{\em J. Phys. A:
  Math. Theor.} {\bf 45} (2012)  135207},
  \href{http://arxiv.org/abs/1110.2404}{{\tt arXiv:1110.2404 [math-ph]}}.

\bibitem{STN01}
M.~Shiroishi, M.~Takahashi, and Y.~Nishiyama, ``Emptiness Formation Probability
  for the One-Dimensional Isotropic XY Model,''
  \href{http://dx.doi.org/10.1143/JPSJ.70.3535}{{\em J. Phys. Soc. Jpn.} {\bf
  70} (2001)  3535}, \href{http://arxiv.org/abs/cond-mat/0106062}{{\tt
  cond-mat/0106062}}.

\bibitem{FA05}
F.~Franchini and A.~G. Abanov, ``Asymptotics of Toeplitz determinants and the
  emptiness formation probability for the XY spin chain,''
  \href{http://dx.doi.org/10.1088/0305-4470/38/23/002}{{\em J. Phys. A: Math.
  Gen} {\bf 38} (2005)  5069},
  \href{http://arxiv.org/abs/cond-mat/0502015}{{\tt cond-mat/0502015}}.

\bibitem{Stephan14}
J.-M. Stephan, ``Emptiness formation probability, Toeplitz determinants, and
  conformal field theory,''
  \href{http://dx.doi.org/10.1088/1742-5468/2014/05/P05010}{{\em J. Stat.
  Mech.} (2014)  P05010}, \href{http://arxiv.org/abs/1303.5499}{{\tt
  arXiv:1303.5499 [cond-mat.stat-mech]}}.

\bibitem{NR15}
K.~{Najafi} and M.~A. {Rajabpour}, ``{Formation probabilities and Shannon
  information and their time evolution after quantum quench in transverse-field
  XY-chain},'' {\em ArXiv e-prints} (2015)  ,
  \href{http://arxiv.org/abs/1511.06401}{{\tt arXiv:1511.06401
  [cond-mat.str-el]}}.

\bibitem{EP07}
V.~Eisler and I.~Peschel, ``Evolution of entanglement after a local quench,''
  \href{http://dx.doi.org/10.1088/1742-5468/2007/06/P06005}{{\em J. Stat.
  Mech.} (2007)  P06005}, \href{http://arxiv.org/abs/cond-mat/0703379}{{\tt
  cond-mat/0703379}}.

\bibitem{cardy-calabrese-quench1}
P.~{Calabrese} and J.~{Cardy}, ``{Time Dependence of Correlation Functions
  Following a Quantum Quench},''
  \href{http://dx.doi.org/10.1103/PhysRevLett.96.136801}{{\em Physical Review
  Letters} {\bf 96} (2006) no.~13, 136801},
  \href{http://arxiv.org/abs/cond-mat/0601225}{{\tt cond-mat/0601225}}.

\bibitem{gritsev-stb-neel-to-xx-first}
P.~{Barmettler}, M.~{Punk}, V.~{Gritsev}, E.~{Demler}, and E.~{Altman},
  ``{Relaxation of Antiferromagnetic Order in Spin-1/2 Chains Following a
  Quantum Quench},''
  \href{http://dx.doi.org/10.1103/PhysRevLett.102.130603}{{\em Physical Review
  Letters} {\bf 102} (2009) no.~13, 130603},
  \href{http://arxiv.org/abs/0810.4845}{{\tt arXiv:0810.4845
  [cond-mat.other]}}.

\bibitem{gritsev-demler-stb-quench-osszf-neel-to-xx}
P.~{Barmettler}, M.~{Punk}, V.~{Gritsev}, E.~{Demler}, and E.~{Altman},
  ``{Quantum quenches in the anisotropic
  spin-$\backslash$frac$\{$1$\}$$\{$2$\}$ Heisenberg chain: different
  approaches to many-body dynamics far from equilibrium},''
  \href{http://dx.doi.org/10.1088/1367-2630/12/5/055017}{{\em New Journal of
  Physics} {\bf 12} (2010) no.~5, 055017},
  \href{http://arxiv.org/abs/0911.1927}{{\tt arXiv:0911.1927
  [cond-mat.quant-gas]}}.

\bibitem{andraschko-sirker}
F.~{Andraschko} and J.~{Sirker}, ``{Dynamical quantum phase transitions and the
  Loschmidt echo: A transfer matrix approach},''
  \href{http://dx.doi.org/10.1103/PhysRevB.89.125120}{{\em Phys. Rev. B} {\bf
  89} (2014) no.~12, 125120}, \href{http://arxiv.org/abs/1312.4165}{{\tt
  arXiv:1312.4165 [cond-mat.str-el]}}.

\bibitem{XX-quench-brockmann}
P.~P. {Mazza}, J.-M. {St{\'e}phan}, E.~{Canovi}, V.~{Alba}, M.~{Brockmann}, and
  M.~{Haque}, ``{Overlap distributions for quantum quenches in the anisotropic
  Heisenberg chain},''
  \href{http://dx.doi.org/10.1088/1742-5468/2016/01/013104}{{\em Journal of
  Statistical Mechanics: Theory and Experiment} {\bf 2016} (2016) no.~1,
  013104}, \href{http://arxiv.org/abs/1509.04666}{{\tt arXiv:1509.04666
  [cond-mat.str-el]}}.

\bibitem{Abramowitz}
M.~Abramowitz and I.~Stegun, eds., {\em Handbook of Mathematical Functions}.
\newblock Dover, New York, 1972.

\bibitem{PE09}
I.~Peschel and V.~Eisler, ``Reduced density matrices and entanglement entropy
  in free lattice models,''
  \href{http://dx.doi.org/10.1088/1751-8113/42/50/504003}{{\em J. Phys. A:
  Math. Theor.} {\bf 42} (2009)  504003},
  \href{http://arxiv.org/abs/0906.1663}{{\tt arXiv:0906.1663
  [cond-mat.stat-mech]}}.

\end{thebibliography}\endgroup
\bibliographystyle{utphys}

\end{document}